\documentclass[conference]{IEEEtran}
\makeatletter
\def\ps@headings{%
\def\@oddhead{\mbox{}\scriptsize\rightmark \hfil \thepage}%
\def\@evenhead{\scriptsize\thepage \hfil \leftmark\mbox{}}%
\def\@oddfoot{}%
\def\@evenfoot{}}
\makeatother
\ifCLASSINFOpdf

\else

\fi

\usepackage{graphicx}
\usepackage{float}
\usepackage{amsmath,amsthm,amssymb,amsfonts}
\usepackage{threeparttable}
\usepackage{subfigure}

\usepackage[ruled,vlined,commentsnumbered]{algorithm2e}
\usepackage{amsmath}
\usepackage{times}

\theoremstyle{plain}
\newtheorem{thm}{Theorem}
\theoremstyle{plain}

\theoremstyle{plain}
\newtheorem{defn}{Definition}

\makeatletter
\thm@headfont{\sc}
\makeatother
\newtheorem{prop}{Proposition}

\begin{document}

\title{Secure Compressed Reading in Smart Grids}

\author{
\IEEEauthorblockN{Sheng Cai, Jihang Ye, Minghua Chen, Jianxin Yan, Sidharth Jaggi}
\IEEEauthorblockA{Department of Information Engineering\\
The Chinese University of Hong Kong\\
N.T., Hong Kong\\
Email: \{cs010, yjh010, minghua, jxyan, jaggi\}@ie.cuhk.edu.hk}}
\maketitle

\begin{abstract}
Smart Grids measure energy usage in real-time and tailor supply and delivery accordingly, in order to improve power transmission and distribution. For the grids to operate effectively, it is critical to collect readings from massively-installed smart meters to control centers in an efficient and secure manner. In this paper, we propose a secure compressed reading scheme to address this critical issue. We observe that our collected real-world meter data express strong temporal correlations, indicating they are sparse in certain domains. We adopt Compressed Sensing technique to exploit this sparsity and design an efficient meter data transmission scheme. Our scheme achieves substantial efficiency offered by compressed sensing, without the need to know beforehand in which domain the meter data are sparse. This is in contrast to traditional compressed-sensing based scheme where such sparse-domain information is required a priori. We then design specific dependable scheme to work with our compressed sensing based data transmission scheme to make our meter reading reliable and secure. We provide performance guarantee for the correctness, efficiency, and security of our proposed scheme. Through analysis and simulations, we demonstrate the effectiveness of our schemes and compare their performance to prior arts.
\end{abstract}

\IEEEpeerreviewmaketitle

\section{Introduction} \label{intro}

\par Smart Grids are playing a significant role in leading the global electrical grids revolution \cite{Obama}. A recent trend is to deploy advanced information control and communication techniques in Smart Grids for better power transmission and distribution \cite{SGload}. It has attracted significant attentions from government, industry, and academics \cite{AUSSG}.

\par One promising Smart Grids architecture based on wireless has been proposed in \cite{Obama}\cite{TMSG}. Wireless broadband networks can serve more wide-area and mission-critical utility communications due to its flexibility and high-speed transmission; service providers always choose them to improve reliability and resiliency during emergency scenarios. Thus, a hardened commercial wireless data network can serve as core part in building overall Smart Grids networks and exploit the advantage of elastic deployment with dynamic routing.

\par To fully unleash such wireless-based Smart Grids' potential and fulfill their design objective, it is critical to collect data adaptively in a wireless manner from smart meters installed in millions of households with efficiency and dependability guarantee \cite{SGC10_SEC_Liu}\cite{SMG_CS}\cite{SGC10_SEC_Bartoli}. This critical problem has only been partially explored recently. We present a summary of related work on this topic in Section \ref{related}.

\par In this paper, we provide an efficient and secure solution to collect measurements from all smart meters in wireless-based Smart Grids. Inspired by compressed sensing theory \cite{CS_src}, substantial studies \cite{Mobicom09}\cite{ICC2010}\cite{SECON} have been explored to improve data transmission efficiency and it also gets applied in Smart Grids \cite{SMG_CS}. We exploit the observed correlations among Smart Grids meter readings  and adopt the compressed sensing technique to design an efficient transmission scheme. We also propose protection mechanism tailored for our compressed-sensing based reading transmission to achieve reliability and security. In particular, we make the following contributions:
\begin{itemize}
  \item From our collected real-world trace, we observe that the smart meter readings demonstrate strong temporal correlations. This indicates they are sparse in certain (unknown) domains.
  \item We design an adaptive compressed-sensing based scheme to collect data. The scheme has two salient features. First, data collection works under arbitrary tree topologies. Second, it works without the need to know a priori in which domain the readings are sparse. This is in contrast to traditional compressed-sensing based scheme where such sparse domain information is required a priori. Performance guarantee of our scheme is also presented.
  \item We design specific dependable scheme that can work with our compressed transmission to make it reliable and secure. Our dependable scheme considers physical link failures, outside and semi-honest inside attacks.
  \item We carry out extensive numerical experiments using real-world smart meter readings to evaluate our scheme from transmission cost to reconstruction performance.
\end{itemize}

\par The rest of our paper is organized as follows. We review related works in Section \ref{related}. Compressed reading and reconstruction scheme is discussed in Section \ref{trans}. Corresponding dependable mechanism in combination with transmission scheme is presented in Section \ref{sec}. Experimental results and analysis are shown in Section \ref{exp}. Finally, we conclude our work in Section \ref{conclude}.

\section{Related Work} \label{related}

\par Data collection problem in Wireless Sensor Networks (WSN) has received extensive studies \cite{Mobicom09}-\cite{SECON}. Data transmission in wireless-based Smart Grids shares many similarities with WSN, such as real-time transmission and dynamic routing; however, there're still two primary differences. First, Smart Grids networks is only tree topology while WSN is arbitrary topology, specific topology allows tailored solution design to maximize the performance. For instance, our scheme utilizes the tree structure to specify node behaviors and reduce transmission cost; secondly, meter readings express specific correlations that other WSN data might not express. We further utilize the strong temporal relationships among readings to improve transmission efficiency.

\par Emerging problem of meter data collection in Smart Grids has attracted much attention recently. We summarize the difference between our work
and existing work in Table \ref{Table:comparison}.

\begin{table}[ht!]
\centering \caption{Comparison with previous work for Reading Transmission in WSN and Smart Grids}
\begin{tabular}{|c|c|c|c|c|}
\hline
 & \textbf{A} & \textbf{B} & \textbf{C} & \textbf{D}\tabularnewline
\hline
Bartoli \emph{et al.} \cite{SGC10_SEC_Bartoli}   & \textbf{$\mathbf{\checkmark}$} &  & Individual Data & \textbf{$\mathbf{\checkmark}$} \tabularnewline
\hline
Li and Liu \cite{SGC10_SEC_Liu}& \textbf{$\mathbf{\checkmark}$} &  & Group Summation &\textbf{$\mathbf{\checkmark}$} \tabularnewline
\hline
Li \emph{et al.} \cite{SMG_CS}  & \textbf{$\mathbf{\checkmark}$}  & \textbf{$\mathbf{\checkmark}$} & Individual Data & \tabularnewline
\hline
C.Luo \emph{et al.} \cite{Mobicom09} &  & \textbf{$\mathbf{\checkmark}$} & Individual Data & \tabularnewline
\hline
J.Luo \emph{et al.} \cite{SECON}\cite{ICC2010} &  & \textbf{$\mathbf{\checkmark}$} & Individual Data & \tabularnewline
\hline
This paper  & \textbf{$\mathbf{\checkmark}$} & \textit{O(M)}$\sim$\textit{O(M$\cdot$N)} & Individual Data & \textbf{$\mathbf{\checkmark}$}\tabularnewline
\hline
\end{tabular}
\begin{tablenotes}
\item \textbf{A}: Security; \textbf{B}: Transmission Efficiency; \textbf{C}: Granularity of Transmission;
\item \textbf{D}: Stream Data Transmission; \textbf{$\mathbf{\checkmark}$} means incorporated in work
  \end{tablenotes}
\label{Table:comparison}%
\end{table}

\par Bartoli \emph{et al.} \cite{SGC10_SEC_Bartoli} considered collecting individual meter readings independently and securely. The scheme has low transmission efficiency, since it does not explore correlation across data and every meter measurement is transmitted independently. Li \emph{et al.} \cite{SMG_CS} further explored the correlation across meter readings and applied compressed sensing to improve efficiency; however, it devised centralized security mechanism based on wireless AP and lacked reliability. Li and Liu \cite{SGC10_SEC_Liu} considered secure aggregate meter data collection with homomorphic encryption. The scheme ensures secure transmissions but only collects aggregated readings, while individual reading collection is required under most scenarios. As comparison, our scheme explores correlation across meter readings and develop transmission solution based on compressed sensing. Moreover, reliability and security concerns are also critical in Smart Grids and have been considered \cite{General_Sec_1}\cite{General_Sec_2}, we also developed specific scheme to warranty security during transmission.

\par In recent years, compressed sensing have been explored in both signal processing and data transmission communities due to its high efficiency
and good recovery performance \cite{CS_src}. In \cite{Mobicom09}, C.Luo \emph{et al.} first considered efficient data transmission using compressed sensing in WSN and acts as the baseline scheme for our work. Then J.Luo \emph{et al.} \cite{SECON}\cite{ICC2010} applied hybrid compressed sensing also in WSN data transmission and achieve improved efficiency while failed to consider the case under stream data over multiple time-slots. For reading transmissions without compressed sensing, they can be easily extended to stream data case; however, it's non-trivial for schemes using compressed sensing as Section \ref{4.d} explained. Inspired by this, we develop efficient compressed reading transmission for Smart Grids which can work under stream reading collection. Besides security guarantee, the differences between our scheme and \cite{Mobicom09}\cite{SECON} are followings: first, our application scenario is Smart Grids; second, it achieves good efficiency with consideration of adaptively transmission for stream readings.

\section{Problem Setting}\label{problem}

\begin{table}
\begin{center}
\caption{Notation List in Algorithms}\label{notation}
\begin{tabular}{|c|c|}
  \hline
  Symbols & Notations \\
  \hline
  G & predefined transmission topology \\
  \hline
  V & Set of transmission participating nodes\\
  \hline
  E & Set of wireless links\\
  \hline
  $d_{i}(t)$ & Meter reading at time $t$ for $i \in V$, $d_{i}(t)\in\mathbb{R}^{+}$\\
  \hline
  $\Phi$ & Sensing Matrix\\
  \hline
  $\Psi$ & Wavelet Transform\\
  \hline
  $H$ & Switching matrix\\
  \hline
  $h$[$X$] & hash function for message $X$ \\
  \hline
  ($X$)$_{K}$ & encrypt message $M$ with key $X$ \\
  \hline
  $K_{i}$, $K^{-1}_{i}$ & public$/$private key LHU i \\
  \hline
  \textbf{K}, \textbf{K}$^{-1}$ & public$/$private key list for all LHUs \\
  \hline
  \textbf{K}$_{DC}$ & public key from data collector \\
  \hline
  \textbf{T}$_{S}$ & current time-stamp \\
  \hline
\end{tabular}
\end{center}
\end{table}

\par In this section we present the general setting of our dependable compressed reading scheme in Smart Grids networks and describe our chief goals. First, a list of key notations are given in Table \ref{notation} for both transmission and dependable mechanisms.

\par
\begin{figure}[ht!]
\centering{}\includegraphics[width=0.3\textwidth]{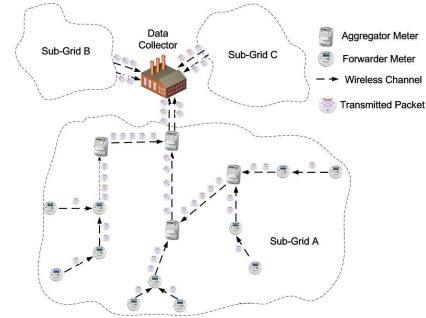}
\caption{Smart Grids Data Transmission Topology\label{trans_topology}}
\end{figure}

 For Smart Grids data transmission, we consider arbitrary tree topology as depicted in Fig. \ref{trans_topology} where data collector is the root. We assume time is chopped into equal-length slots and represent Smart Grids networks as $G$ $=$ $(V,E)$. $V$ is the set of nodes in tree. Every node, except the root, represents a smart meter. For every node $i\in V$ , we use $d_{i}(t)\in\mathbb{R}^{+}$ to represent its meter reading at time $t$. Nodes coordinate to transmit readings from all meters, i.e., \{$d_{i}(t)$\}$_{t}$ for all $i\in V$, to the root which represents the Data Collector. $E$ is the set of wireless links where one node can directly communicate with its parent. We assume that all links can transmit one message in every slot simultaneously (through locally orthogonal channels or properly imposed scheduler).

More specifically, we regard direction towards data collector as upstream and consider three types of nodes in Smart Grids:
\begin{itemize}
\item Forwarder: Legitimate Home Users (LHUs) that reside mostly on downstream of the tree. A Forwarder first acquires reading from attached smart-meter, then forwards to its parent node together with the data from its downstream children nodes.
\item Aggregator: Legitimate Home Users (LHUs) that usually reside in the middle of the tree. An Aggregator collects reading from its smart-meter, aggregates it with recited data from downstream children, and sends aggregated data to parent. Further explanations in Section \ref{4.c}.
\item Data Collector: the root node of the tree. It receives all the readings sent from Aggregators and Forwarders, and reconstructs original readings of all smart meters.
\end{itemize}
\par Detailed definitions for Forwarder and Aggregator will be specified in Section \ref{4.c} where we introduce our compressed reading scheme in Algorithm\ref{CSAlg}.

\par For dependability-related issues, first we try to ensure reliability with consideration of link failure; then we assume Smart Grids data transmission being exposed to outside attackers who can destroy transmitted readings or impersonate LHUs, they are further explored in Section \ref{sec}. Smart Grids' LHUs are regarded as semi-honest who would passively eavesdrop readings instead of actively tampering.

\par There are two objectives in designing a dependable compressed reading scheme. The first objective is to use as few total number of transmissions to allow all smart meters' reading reconstruction up to acceptable bounded errors, i.e., \{$d_{i}(t)$\}$_{t}$ for all $i\in V$ at Data Collector. The second objective is to make the transmissions dependable, by designing mechanism tailored for the compressed transmission, we generate trouble-free transmission topology and defend against outsiders, semi-honest insiders.

\section{Compressed Reading and Reconstruction}\label{trans}

\subsection{Compressed Sensing Preliminary}

We first briefly review the necessary compressed sensing background. In general, one needs $N$ measurements to fully recover an $N$-dimensional signal. However, for an $N$-dimensional signal is sparse in certain domain, the compressed
sensing theory \cite{CS_src} states a rather surprising result: one can
fully recover the signal by using only $\Omega(\log N)$ linear measurements.
\begin{defn}
An $N$-dimensional signal $\boldsymbol{d}$ is said to be $K$-sparse
in a domain $\Psi$, if there exists an $N$-dimensional vector $\boldsymbol{x}$
so that $\boldsymbol{d}=\Psi\boldsymbol{x}$ and $\boldsymbol{x}$
has at most $K$ non-zero entries ($K<N$).
\end{defn}
\textbf{Remarks:} (i) the above definition covers the case where $\boldsymbol{d}$
is sparse itself, for which we can simply take $\Psi=\boldsymbol{I}$
and $\boldsymbol{x}=\boldsymbol{d}$. (ii) many natural signals are
sparse in certain domain. For example, natural images are sparse in  Wavelet domain \cite{Mallat}. We observe meter readings are
sparse in frequency domain.

Let $\boldsymbol{y}$ be an $M$-dimensional linear measurements of
$\boldsymbol{d}$, i.e., $\boldsymbol{y}=\Phi\boldsymbol{d}$, where
$\Phi$ is an $M\times N$ \emph{sensing} matrix.
\begin{defn}
An $M\times N$ sensing matrix $\Phi$ is said to satisfy a Restricted
Isometry Property (RIP) of sparsity $K$ ($K<N)$ if there exists
a $\delta_{K}\in(0,1)$ such that the following holds for any $K$-sparse
$N$-dimensional vector $\boldsymbol{z}$:
\[
(1-\delta_{K})||\boldsymbol{z}||_{l_{2}}^{2}\leq||\mathbf{\Phi}\boldsymbol{z}||_{l_{2}}^{2}\leq(1+\delta_{K})||\boldsymbol{z}||_{l_{2}}^{2},
\]
 where $\left\Vert \cdot\right\Vert _{l_{2}}$ represents the $l_{2}$
norm.
\end{defn}

It has been shown in \cite{DLP}\cite{CS_src}\cite{SP} that
an $M\times N$ matrix $\Phi$ satisfies RIP with probability $1-O(e^{-\gamma N})$
for some $\gamma>0$ if
\begin{itemize}
\item all its entries are independent and identically distributed Gaussian
random variables with mean zero and variance $1/M$,
\item and $M\geq\mbox{const}\cdot K\cdot\log N/K$.
\end{itemize}
The following observation is due to \cite{CS_src}:
\begin{thm}
Consider an under-determined linear system $\boldsymbol{y}=\Phi\boldsymbol{d}$
where $\boldsymbol{y}$ is an $M\times1$ vector and $\boldsymbol{d}$
is a $N\times1$ vector and is $K$-sparse in domain $\Psi$. If the
$M\times N$ sensing matrix $\Phi$ satisfies the RIP property and
$\Psi$ is orthornomal, then $\boldsymbol{d}$ can be recovered exactly
by solving the following convex optimization problem
\begin{eqnarray}\label{originCS}
\textmd{min}_{\boldsymbol{x}} &  & ||\boldsymbol{x}||_{l_{1}}\\
\mbox{s. t.} &  & \boldsymbol{y}=\Phi\boldsymbol{d}\nonumber \\
 &  & \boldsymbol{d}=\Psi\boldsymbol{x}.\nonumber
\end{eqnarray}

\end{thm}

\subsection{Challenges and Solution Overview}

Since the energy consumption habit is lasting, which indicates that the meter readings have strong temporal correlation (detailed in Section \ref{exp}), thus meter readings in Smart Grids are sparse in frequency domain. It is natural to explore compressed sensing to reduce
the number of transmissions for the data collector to collect all
the meter readings.

However, it is nontrivial to apply compressed sensing in our problem.
In particular, there are three challenges stand:
\begin{itemize}
\item How to generate sensing matrix $\Phi$ which satisfies RIP in Smart Grids networks and data collector can recover the original data using the same $\Phi$ without receiving it directly;
\item How to select sparse domain $\Psi$ to represent the meter readings sparsely since the readings are not sparse naturally and compressed sensing deals with the signal which is compressible in some domain;
\item How to adjust $\Psi$ adaptively to deal with a stream of data since we need to recover a stream of data which is not sparse in a fixed domain.
\end{itemize}
We address these three challenges and successfully design an efficient
meter reading solution based on compressed sensing. We address the
first challenge by using the pseudo-random number generator seeded with
the node's identity. For the second and the third challenges, we design
transform domain by using the property of piecewise polynomial signal.
We elaborate our solutions in the next two subsections.

\subsection{Compressed Reading}\label{4.c}

In this section, we will give a transmission scheme to address first
challenge to apply compressed sensing to Smart Grids networks.

Given Smart Grids networks $G(V,E)$, $|V|=N$. Denote $S$ is
the Data Collector. $M$
is compressed factor defined by data collector and data collector broadcasts it to the entire network. We define forwarder as the node whose number of children nodes is less than or equal to $M-1$ and aggregator as the node whose number of children nodes is larger than $M-1$. For each node $i\in V,$ $ID_{i}$ is its identity.
$C_{i}$ is the set of children nodes of node $i$. $F_{i}$ and $A_{i}$
is the set of forwarder children nodes and set of aggregator children
nodes of it respectively. $m_{i}$ is the message it sends out.

We make some assumptions that are considered to be reasonable in our
scheme:
\begin{itemize}
\item Every LHU reports its data synchronously. Also, each LHU should
append its ID within transmitted message.
\item Data Collector receives measurements periodically from all registered
LHUs; it then recovers the data and estimate the grid state at
that moment.
\end{itemize}

In Algorithm\ref{CSAlg}, if there is no aggregator in the network, then each forwarder just relays the readings from its children nodes including its own reading to the parent node on the tree. If not, the forwarders just work as the same way as the previous case while each aggregator will generate weighted sum of the readings from its forwarder children nodes and combine with the message from its aggregator children nodes, then reports to its parent node. Finally, the data collector will get the measurements of all data readings for the purpose of reconstruction of original data.
It's clear that each node reports at most $M$ messages in the transmission scheme. And total cost is even less than the scheme in \cite{Mobicom09}. Due to applying compressed sensing to the data transmission in Smart Grids, the bottleneck and total cost of Smart Grids networks will be significantly reduced. Detailed discussion in Section.\ref{4.f}.

\begin{algorithm}\label{CSAlg}
\DontPrintSemicolon \KwIn{{[} $G(V,E)$, $S$, $ID_{i}$, $d_{i}$,
$\forall i\in V$, $M$ {]}} \KwOut{{[} $m_{i}$, $\forall i\in V${]}}

\Begin{ Count $|C_{i}|$, $\forall i\in V$ \;
Generate $A_{i}$ and $F_{i}$, $\forall i\in V$\;
\If {$|\cup_{j\in V}A_{j}|=0$}{$\forall i\in V$, $m_{i}$ $=$
${\cup_{j}}$$d_{j}$, $j\in C_{i}\cup{i}$ }\lElse{

\For {$l=1$ to $M$}{ \If {$i$ is forwarder}{ $m_{i}$ $=$
${\cup_{j}}$$d_{j}$, $j\in C_{i}\cup{i}$ } \lElse
{{$i$ generates gaussian random coefficients $\phi_{lj}$ using
a pseudo-random number generator for all $j\in F_{i}\cup{i}$ seeded
with associated $ID_{j}$\; $m_{i}$=$\sum_{j\in F_{i}}\phi_{lj}d_{j}+\sum_{j\in A_{i}}m_{j}$\;}
}
{ $S$ collects the weight measurement of all nodes represented
as $y_{l}=\sum_{j=1}^{N}\phi_{lj}d_{j}$ } }}  S can get the
mathematic formulas $y=\Phi d$ to demonstrate the all measurements} \caption{Compressed Reading Algorithm}
\end{algorithm}

\subsection{Compressed Reconstruction}\label{4.d}

In this section, we will propose data reconstruction scheme based
on compressed sensing for Data Collector to recover original data
after collecting all $M$ measurements using the transmission scheme.
When the network topology is known by Data Collector, it can generate
the same sensing matrix $\Phi$ using the same pseudo-random number
generator and the same $ID$. However, the prerequisite of exact reconstruction
of using compressed sensing technology is that the signal is sparse
or sparse in certain domain. We propose the scheme below to address
this challenge. The objective of our scheme is to reconstruct the
meter readings of all smart meters $d_{i}(t)$, $\forall i\in V$
by Data Collector at time $t$. Let's first look into the simple case
where Data Collector need to recover the static data. After that,
we will address how to deal with the stream data.

\subsubsection{Snapshot Case}

In this case, We propose the algorithm for Data Collector to reconstruct
the static data $\boldsymbol{d}\in \mathbb{R}^{N}$.

It is well known that piecewise polynomial signal is sparse in wavelet domain \cite{Mallat}.

\begin{prop}\cite{walker}
If a signal $\boldsymbol{f}$ is equal to a polynomial of degree
less than J/2 over the support of a k-level DaubJ wavelet $W_{m}^{k}$,
then the k-level coefficient $x_{m}=\boldsymbol{f}\cdot W_{m}^{k}$ is
zero.
\end{prop}

Let $H$ be the switching matrix such that the entries of $H\boldsymbol{d}$ is ascendingly
ordered which can be regarded as piecewise polynomial.

\begin{thm}
Given $\boldsymbol{y}=\Phi\boldsymbol{d}$ where $\boldsymbol{y}$
is a $M\times1$ vector and $\boldsymbol{d}$ is $K$-sparse in domain
$H^{-1}\Psi$. If the entries of $M\times N$ sensing matrix $\Phi$
are independent and identically distributed Gaussian random variables
with mean zero and variance $1/M$ and $\Psi$ is wavelet domain, then Data Collector can
recover $\boldsymbol{d}$ exactly by solving the following convex
optimization problem with probability $1-O(e^{-\gamma N})$ for some $\gamma>0$
if $M\geq\mbox{const}\cdot K\cdot\log N/K$
\begin{eqnarray}\label{ptbCS}
\textmd{min}_{\boldsymbol{x}} &  & ||\boldsymbol{x}||_{l_{1}}\\
\mbox{s. t.} &  & \boldsymbol{y}=\Phi\boldsymbol{d}\nonumber \\
 &  & \boldsymbol{d}=H^{-1}\Psi\boldsymbol{x}.\nonumber
\end{eqnarray}
\end{thm}

\subsubsection{Stream Case}

We have shown that how to reconstruct the static data using the snapshot
algorithm. However, for snapshot algorithm, we assume to know the
H, the switching matrix, in advance but it is usually not known a priori in practice. Further, instead of a snapshot, we need to recover a stream of data \{$d_{i}$($t$)\}$_{t}$ for all $i\in V$.
We address these challenges by designing an algorithm exploiting the temporal correlation and our particular compressed sensing scheme as follows.

\begin{itemize}
\item $t_{0}$: Each node is treated as forwarder and reports its data, so the data collector can get the exact data $\boldsymbol{d(t_{0})}\in \mathbb{R}^{N}$. Data collector sorts the data as ascending order using the switching matrix $H(t_{0})$.
\item $t_{i}$, i=1,2,... : Use compressed reading scheme to get weighted measurements $\boldsymbol{y(t_{i}})=\Phi$$\boldsymbol{d(t_{i})}$. From snapshot case analysis, we know $\boldsymbol{d(t_{i})}$ is $K$-sparse in $H^{-1}(t_{i})\psi$ and can be reconstructed by solving an $l_{1}$ minimization problem. The challenge is that we don't know  $H^{-1}(t_{i})$ before we reconstruct $\boldsymbol{d(t_{i})}$. Thus, data collector takes a bold approach to construct a $\boldsymbol{\hat{d}(t_i)}$ by solving following $l_{1}$ minimization problem where the unknown  $H^{-1}(t_{i})$ is replaced by the available one $H^{-1}(t_{i-1})$:
\begin{eqnarray}\label{stream}
\textmd{min}_{\boldsymbol{x}} &  & ||\boldsymbol{x}||_{l_{1}}\\
\mbox{s. t.} &  & \boldsymbol{y(t_{i})}=\Phi\boldsymbol{d(t_{i})}\nonumber \\
 &  & \boldsymbol{d(t_{i})}=H^{-1}(t_{i-1})\Psi\boldsymbol{x}.\nonumber
\end{eqnarray}

The data collector sorts the estimated data $\boldsymbol{\hat{d}(t_{i})}$
as ascending order to get the estimated switching matrix $H(t_{i})$. The system then proceed to the $i+1^{th}$ round.
\end{itemize}

\par The above algorithm shares a similar flavor as the differential-coding based video coding schemes which encode the initial frame and then only holds the changes from previous one \cite{videoCoding}. The difference is that, in our algorithm, reconstruction of the data in the current slot only depends on \emph{the order} of the data in the previous slot; while in the differential coding, reconstruction of the former one depends on the values of the latter one.

\par There are two issues to consider for the above algorithm. First, how accurate is $\boldsymbol{\hat{d}(t_{i}})$ as compared to the real one $\boldsymbol{d(t_{i})}$? Second, since data collector recovers $\boldsymbol{d_{i+1}}$ using the estimated switching matrix $H(t_{i})$ where there has already existed error. Then, will the reconstruction error amplify into future rounds and deteriorate?

\par In the next, we will bound the error and show that error will not amplify.

\begin{thm}\label{errbnd}
Given specific $\delta_{k}$ satisfying:
\[
(1-\delta_{K})||\boldsymbol{z}||_{l_{2}}^{2}\leq||\mathbf{\Phi\Psi}\boldsymbol{z}||_{l_{2}}^{2}\leq(1+\delta_{K})||\boldsymbol{z}||_{l_{2}}^{2},
\] for any $K$-sparse
$N$-dimensional vector $\boldsymbol{z}$ and definitions of $\Phi$ and $\Psi$ in (\ref{originCS}). $H^{-1}$ is the perturbation matrix. $||X||_{l_{2}}^{k}$ denotes the maximum $l_{2}$ norm of matrix $X$'s arbitrary $k$ columns sub-matrices.

\begin{gather}\label{errdef}
\begin{split}
\gamma_{A}=\frac{||\Phi H^{-1}\Psi-\Phi\Psi||_{l_{2}}^{k}}{||\Phi\Psi||_{l_{2}}^{k}}, \beta_{A}=\frac{\sqrt{1+\delta_{k}}}{\sqrt{1-\delta_{k}}}\\
\gamma_{A}^{'}=\frac{||\Phi H^{-1}\Psi-\Phi\Psi||_{l_{2}}^{2k}}{||\Phi\Psi||_{l_{2}}^{2k}}, \delta_{k}^{'}<\frac{\sqrt{2}}{(1+\gamma_{A}^{'})^{2}}-1\\
C=\frac{4\sqrt{1+\delta_{k}^{'}}(1+\gamma_{A}^{'})}{1-(\sqrt{2}+1)[(1+\delta^{'}_{k})(1+\gamma_{A}^{'})^{2}-1]}
\end{split}
\end{gather}

Then, $l_{2}$ norm of the difference between original readings, $\boldsymbol{d}$, and the recovered readings to (\ref{ptbCS}), $\boldsymbol{\hat{d}}$, is constrained as.
\begin{equation}\label{errcons}
    ||\boldsymbol{d}-\boldsymbol{\hat{d}}||_{l_{2}}\leq C\beta_{A}\gamma_{A}||\boldsymbol{y}||_{l_{2}}
\end{equation}
\end{thm}

The proof of this theorem is included in Appendix.

\textbf{Remarks:} (i) Theorem 2 in \cite{estbBound} plays significant intermediate step in proving our Theorem \ref{errbnd}; the difference is that original theorem only bounds errors between sparse signals while ours can restrict errors of meter readings which are not originally sparse. (ii) at time-slot $t_{i}$, we can bound the error incurred with the estimated switching matrix by let $H^{-1}=H^{-1}(t_{i-1})H(t_{i})$, $\Psi=H^{-1}(t_{i})\Psi$, $\boldsymbol{d}=\boldsymbol{d(t_{i})}$, $\boldsymbol{y}=\boldsymbol{y(t_{i})}$ and $\boldsymbol{\hat{d}}$ is the recovered readings to (\ref{stream}); (iii) we can choose the "worst" $H^{-1}$ to give the largest bound in (\ref{errcons}), thus error propagation can be ignored. In real Smart Grids data transmission, we get the observation that $H^{-1}$ is far better than random perturbation, it only changes partly every two time-slots. It is supported by real data experiments in Section\ref{exp}.

\subsection{Increment Analysis}
We have demonstrated how to quantify the propagated error under arbitrary interval increments. Here, we consider reasonable constraint against reading increment $\boldsymbol{n(t_{i})}$ from $\boldsymbol{d(t_{i})}$ to $\boldsymbol{d(t_{i+1})}$ ($i=1,2,...$), and propose another approach to bound the estimation error. We find that stronger error bound can be achieved if increment meets certain requirements.

\begin{prop}\label{exact}
If $\boldsymbol{n(t_{i})}$ is $K$-sparse in domain $H^{-1}(t_{i})\Psi$, then we can
reconstruct $\boldsymbol{d(t_{i+1})}$ exactly, i=0,1,...,$\infty$.\end{prop}

The proof of this proposition is included in Appendix.

In real data, the increment may not be exact $K$-sparse in domain $H^{-1}(t_{i})\Psi$. Even though, we still can give the bound of reconstruction error when increment is approximately $K$-sparse. The definition of approximately sparse is given below.

\begin{defn}\label{asparse}
The best $K$-sparse approximation $\boldsymbol{d_{K}}$ of $N$-dimensional signal
$\boldsymbol{d}$ is obtained by keeping the $K$ largest entries of $\boldsymbol{d}$ and setting
the others to zero. An $N$-dimensional signal $\boldsymbol{d}$ is
said to be approximately $K$-sparse in a domain $\Psi$, if there
exists an $N$-dimensional vector $\boldsymbol{x}$ so that $\boldsymbol{d}=\Psi\boldsymbol{x}$
and $||\boldsymbol{x}-\boldsymbol{x_{K}}||_{l_{1}}\leq\varepsilon$, $\varepsilon$ is a positive
constant.
\end{defn}

\begin{prop}\label{approximate}
 If $\boldsymbol{n(t_{i})}$ is approximately
$K$-sparse in domain $H^{-1}(t_{i})\Psi$. Then the estimation error
$||\boldsymbol{\hat{d}(t_{i+1})}-\boldsymbol{d(t_{i+1})}||_{l_{2}}\leq C_{0}K^{-1/2}\varepsilon$, for i=0,1,...,$\infty$ and some constant $C_{0}$.
\end{prop}

The proof of this proposition is included in Appendix.

\textbf{Remarks:} In real data, we can observe strong correlation between $\boldsymbol{n(t_{i})}$ and $\boldsymbol{d(t_{i})}$. Therefore, we can consider that $\boldsymbol{n(t_{i})}$ and $\boldsymbol{d(t_{i})}$ share the same sparse domain.

\subsection{Performance Analysis}\label{4.f}

In this part, we analyze the performance of our transmission cost. First, the minimum and maximum transmission costs are given.

\begin{thm}\label{thmCost}
Over any tree topology, for N LHUs using our transmission scheme, minimum transmission cost is $N$ while maximum transmission cost is $M(N-M/2+1/2)$.
\end{thm}

The proof of this theorem is included in Appendix.

Then, we make transmission cost analysis against one special case: $p$-array complete tree.
\begin{prop}\label{kary}
For $N$ node p-ary complete tree transmission topology, cost in our scheme is $O(N\log_{p}M)$.
\end{prop}

The proof of this proposition is included in Appendix.

While for the scheme in \cite{Mobicom09}, the total cost is always $O(N\cdot M)$ and our scheme achieves better performance than that.

\section{Secure Transmission Mechanism} \label{sec}
\par Previous work such as \cite{SGC10_SEC_Liu}\cite{SMG_CS}\cite{SGC10_SEC_Bartoli} \cite{CCS09}\cite{SGC10_SE_1}\cite{SGC10_SE_2}\cite{SGC10_SEC_Scheme} considered security in smart grids; however, they address security at data collector through either secrets from AP or estimation of grids state to check whether a discrepancy exists with the original, and they assumed that each LHU does not fail. In contrast, our scheme is a distributed solution where both aggregators and forwarders get involved instead of merely relying on data collector, and we consider reliability issues.

\subsection{Reliability}
\par To achieve reliable data transmission, we perform a diagnostic test to settle physical errors, such as link failures and traffic congestion. Before data transmission, a data collector publishes temporary transmission topology $G^{'}$: each LHU has several outgoing links including one primary link and can communicate through them wirelessly. During the test, each LHU transmits test package with primary outgoing links to verify its effectiveness: once failed or suffered serious delays, it would broadcast link failure, enable another outgoing link as primary and continue testing. Iteratively perform this test till all LHUs' primary outgoing links work. Using each node's primary outgoing links, we generate a ready topology $G$, which will be used for reading data transmission.

\subsection{Security}

\par

We protect data transmission from attacks launched by either outsiders or semi-honest insiders, and our security model is the following:

\subsubsection{Insider Adversary} Corrupted LHUs can work as malicious insider attackers. Following the standard assumptions in \cite{SGC10_SEC_Liu}\cite{SGC10_SEC_Bartoli}, we assume that insider adversaries are semi-honest, namely they execute our algorithms  properly, but they want to use received transmission data to infer
other LHU's consumption behaviours. Also, insider adversaries can deny that they have sent a particular data to the collector.

Inside adversaries do not drop or modify received packets, and data corruptions are only caused by outside adversaries.

\subsubsection{Outsider Adversary}we assume that an outsider adversary has a polynomial-bounded computational capacity  and can actively launch the following attacks.
\begin{itemize}
\item Data privacy: the attack can evade a LHU's privacy to infer their consumer behaviours.
  \item Data Tampering: the attacker can tamper measurements along the link to alter or forge measurement and make data collector perform incorrect reconstruction;
  \item Impersonation: the attacker can imitate a LHU, send forged data on its behalf to the collector;
  \item Replay: the attacker can intercept previous transmissions and send them later in following days to cause recovery error;
\end{itemize}

We defend attacks that can be launched by the above adversaries as follows.

\begin{itemize}
  \item \emph{Data Privacy}:  we encrypt transmitted data to hide consumption behaviors. Our transmission scheme requires that numerical calculations be performed on encrypted data. To achieve that, we employ Paillier Crypto-system \cite{Paillier}, which has the following homomorphic properties:

\begin{eqnarray}\label{privacyhomo}
 \nonumber E(m_{1}+m_{2}) = E(m_{1})*E(m_{2}),  E(\phi\cdot m) = E(m)^{\phi} \\
  E(\phi_{1}m_{1}+\phi_{2}m_{2}) = E(m_{1})^{\phi_{1}}*E(m_{2})^{\phi_{2}}
\end{eqnarray}
 
 where $E(m_{1})$ and $E(m_{2})$ are encrypted measurements, $\phi_{1}$ and $\phi_{2}$ are random coefficients generated by a LHU with its ID.
      The Paillier scheme is a public-key crypto-system, and therefore if measurement readings are encrypted with the data collector's public key, only will the data collector can decrypt and reconstruct the readings.

  \item \emph{Data Integrity}: to check whether the encrypted transmissions get damaged, we use a cryptographic hash function to verify data integrity. Since measurement data in our setting is always 4 bytes long, we use SHA-1 hash function, which produces a 160-bit output, but we will only the first 64 bits of the hash value for our integrity verification. This will
reduce data transmission cost, while maintaining reasonable security.

  \item \emph{Impersonation}: We use standard digital signatures to defend against impersonation  attacks, and provide non-repudiation. Each LHU generates a unique RSA key pair: its private key is kept while public is published. Signature can be generated only by LHU with legal private key while can be verified by its public key held by others.

\item \emph{Message Freshness}: We use time stamp to guarantee that each message transmitted is fresh, and thus defend against replay attacks. Specifically, a LHU uses its private RAS key to sign a hashed message together with its time stamp to ensure message freshness and integrity simultaneously.
\end{itemize}

\subsection{Integrating security with our Transmission Scheme}

\par Our secure transmission scheme includes 3 algorithms as follows. 

Algorithm\ref{valid} (notations are explained in Table \ref{notation}.) explains how to validate received packet: it fails if ID doesn't belong to certain set or if integrity verification fails after successful decryption.

\begin{algorithm}
\DontPrintSemicolon
\textbf{Remark}: verify the source ID and integrity of \emph{Pkt}\\
\KwData{[ \textbf{K}, \emph{Pkt}, Set, $T_{S}$ ]\\\textbf{K} are public key list, \emph{Pkt} with typical form, Set is ID set, $T_{S}$ is fixed time-stamp}
\KwResult{[ \textbf{Ans, ID, EncData} ]}
\Begin{
    divide \emph{Pkt} into \emph{E}, \emph{id} and \emph{Sig} from typical form\;
    \textbf{ID} $\leftarrow$ \emph{id} ; \textbf{EncData} $\leftarrow$ \emph{E}\;
        \lIf{ID$\not\in$Set or (\emph{Sig})$_{K[ID]}$$\neq$(h[EncData], $T_{S}$)}{
        \textbf{Ans} = false\;
        }
        \lElse{\textbf{Ans} = true\;}
        return \textbf{Ans, ID, EncData}\;
}
\caption{Validation\label{valid}}
\end{algorithm}

\par Algorithm \ref{FwdSec} is secure transmission scheme for Forwarders. Forwarder LHU$i$ first maintains three sets \textbf{Rec\_P}, \textbf{V}, and \textbf{Rec\_V} respectively recording received packets, IDs of downstream neighbors, and IDs of received packets. Then, he checks all the received packets using \texttt{Validation}: if passed, put this packet and its ID into \textbf{Rec\_P} and \textbf{Rec\_V}; otherwise, just abandon it. After verifications of all received packets, LHU$i$ knows whether all downstream neighbors have sent their measurements; if not, ask LHUs in \textbf{V}$\backslash$\textbf{Rec\_V} for resend: perhaps suffering active tampering attacks. Finally, LHU$i$ encrypts his own measurement with Paillier, makes hash values, signs hash and time-stamp, then concatenates them to generate new packet. He transmits packets from \textbf{Rec\_P} with its own packet directly to predefined parent node.

\par Algorithm \ref{AggSec} is security strategy for Aggregator's transmission. Since aggregator LHU$i$ needn't store and forward received packets, he only maintains sets \textbf{V} and \textbf{Rec\_V}. However, he should distinguish forwarders from aggregators in his downstream neighbors and maintain their IDs in set \textbf{V}$_{f}$ for forwarders and \textbf{V}$_{a}$ for aggregators. Then he validates all the received packets and make compression for encrypted message from Paillier Crypto-system. For packets from non-aggregation transmission, LHU$i$ use $F$ to store compressed value from \textbf{V}$_{f}$'s encrypted measurements as Eqn.\ref{privacyhomo}, the random coefficient are generated using \texttt{randGen} with LHU ID. For encrypted measurements from aggregators in \textbf{V}$_{a}$, LHU$i$ use $A$ to store new compressed value through multiplication from Eqn.\ref{privacyhomo}. Then, it also reports missing IDs for resend. Finally, LHU $i$ encrypts its own measurement, adds with $A$ and $F$ to generate new encrypted compressed reading, then produces new packet with LHU ID, hash and signature. Since LHU$i$ is aggregator, he only transmits one packet to parent node.

\begin{algorithm}
\DontPrintSemicolon
\textbf{Remark}: Secure transmission for Forwarder LHU\\
\SetKwFunction{Validation}{Validation}
\KwData{[ i, \textbf{K}, \textbf{K}$^{-1}$, $T_{S}$, \textbf{G}, $K_{DC}$ ]}
\KwResult{[ \textbf{Send Pkt} ]}
\Begin{
\nl    \textbf{V} $\leftarrow$ IDs of LHU[i]'s downstream neighbors in \textbf{G}\;
\nl    \textbf{Rec\_V} $\leftarrow$ $\phi$ ; \textbf{Rec\_P} $\leftarrow$ $\phi$ \;
 \nl   \For{all received packets of LHU[i]}{
  \nl  R = \Validation{\textbf{K}, Current Pkt, \textbf{V}, $T_{S}$}\;
   \nl \If{R.Ans = true}{
        \emph{add} R.ID to \textbf{Rec\_V}, $Current Pkt$ to \textbf{Rec\_P}\;}   }
\nl    \If{\textbf{V}$\backslash$\textbf{Rec\_V} $\neq$ $\phi$}{
 \nl       request IDs in \textbf{V}$\backslash$\textbf{Rec\_V} for resend}
 \nl   E$_{i}$ $\leftarrow$ (M$_{i}$)$_{K_{DC}}$, New Pkt $\leftarrow$ E$_{i}$ $|$ i $|$ (h[E$_{i}$], $T_{S}$)$_{K^{-1}_{i}}$\;
\nl    \textbf{Send Pkt} = \textbf{Rec\_Pkt} $\cup$ \{New Pkt\}\;
}
\caption{SecureTransmission\_Forawrder\label{FwdSec}}
\end{algorithm}

\begin{algorithm}
\DontPrintSemicolon
\textbf{Remark}: Secure transmission for Aggregator LHU\\
\SetKwFunction{PktGen}{PktGen}
\SetKwFunction{Validation}{Validation}
\SetKwFunction{randGen}{randGen}
\KwData{[ i, \textbf{K}, \textbf{K}$^{-1}$, $t$, \textbf{G}, $K_{DC}$ ]}
\KwResult{[ \textbf{Send Pkt} ]\\\textbf{Remarks} }
\Begin{
\nl    \textbf{V} $\leftarrow$ IDs of LHU[i]'s downstream neighbors in \textbf{G}\;
\nl    \textbf{V}$_{f}$/\textbf{V}$_{a}$ $\leftarrow$ IDs of forwarders/aggregators from \textbf{G}\;
 \nl   \textbf{Rec\_V} $\leftarrow$ $\phi$; F $\leftarrow$ 1 ; A $\leftarrow$ 1 \;
\nl    \For{all received packets of LHU[i]}{
\nl    R = \Validation{\textbf{K}, CurrentPkt, \textbf{V}, t}\;
\nl    \If{R.Ans = true}{
\nl    \textbf{Rec\_V} = \textbf{Rec\_V} $\cup$ \{R.ID\}\;
 \nl   \If{R.ID$\in$\textbf{V}$_{f}$}{
\nl    F = F$\ast$(R.EncData)$^{\randGen{R.ID}}$\;}
\nl    \lElse{    A = A$\ast$(R.EncData)\;}}    }
 \nl   \If{\textbf{V}$\backslash$\textbf{Rec\_V} $\neq$ $\phi$}{
 \nl       request IDs in \textbf{V}$\backslash$\textbf{Rec\_V} for resend}
\nl    New Data $\leftarrow$ F $\ast$ A $\ast$ (\randGen{i}$\ast$m[i])$_{K_{DC}}$\;
\nl    \textbf{Send Pkt} $\leftarrow$ New Data $|$ i $|$ (h[New Data], $t$)$_{K^{-1}_{i}}$\;
}
\caption{SecureTransmission\_Aggregator\label{AggSec}}
\end{algorithm}

\section{Evaluation}\label{exp}

\subsection{Experiments for Data Transmission}
\subsubsection{Settings}
\par In transmission efficiency comparison, we generate some arbitrary tree topologies where nodes are randomly located with data collector at the center to simulate Smart Grids networks, and compare with other previous schemes. Here, we use Box-plot \cite{boxplot} to describe the statistical information of transmission cost.
\par In performance evaluation, we use data from Stanford Powernet open project where readings are real-time energy consumption from household appliances \cite{PowerNet}. We collected readings of 128 appliances every twenty minutes over six working days since advanced household smart meters can report measurements at a minimum interval of 15 minutes \cite{Tex15}. After data preprocess of filtering invalid readings as negative or null, we get 396 groups of data for these 128 appliances. Each group includes data of one transmission round for 128 nodes. For simplicity, we assign each node with ID from 1 to 128. Numerical operations are under MATLAB environment.

\subsubsection{Transmission Efficiency}
\begin{figure*}[ht!]
\centering
\subfigure[Cost Comparison-128]{\includegraphics[width=4.2cm]{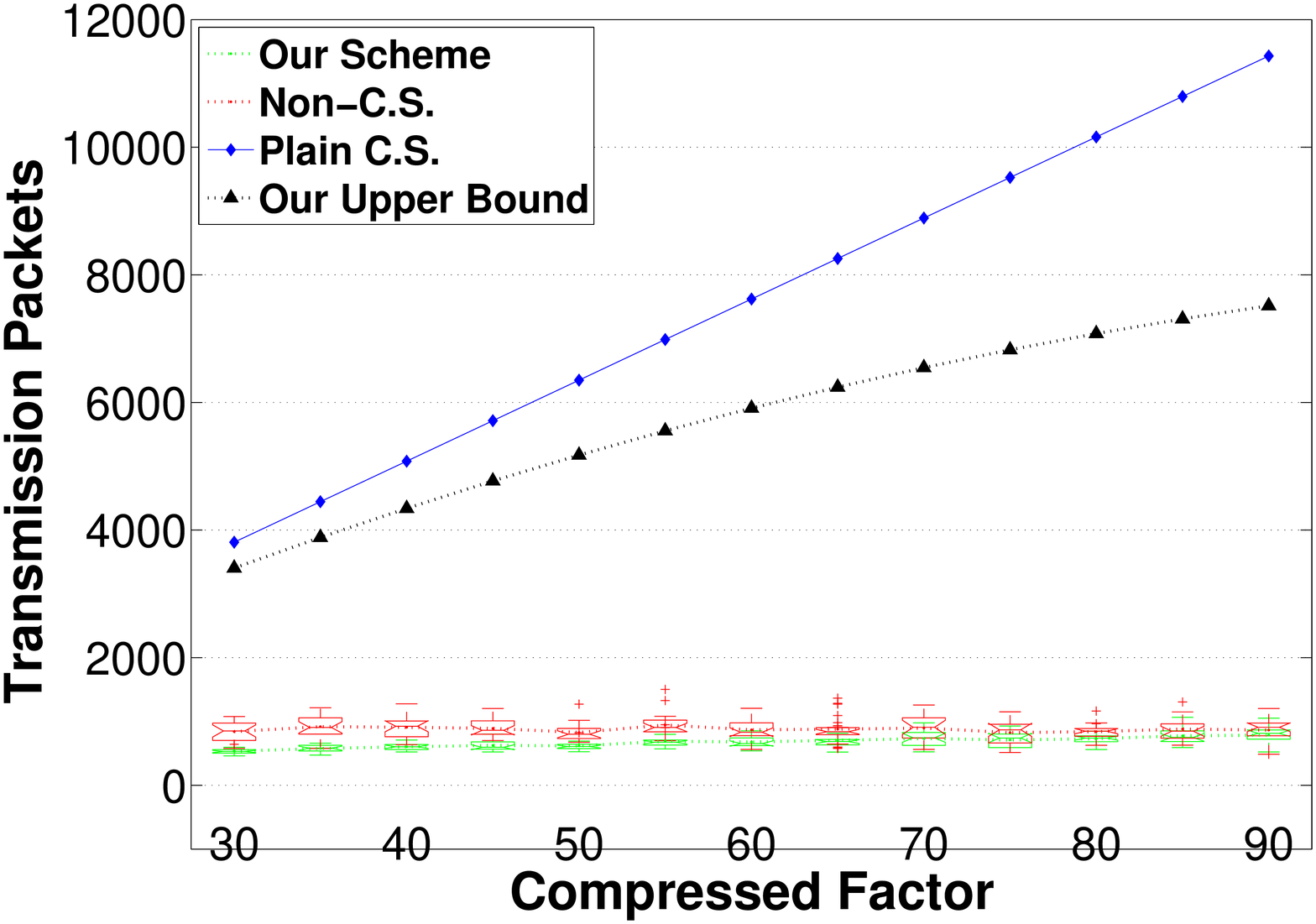}\label{Fig:tran128}}
\subfigure[Comparison with Non-C.S.-128]{\includegraphics[width=4.2cm]{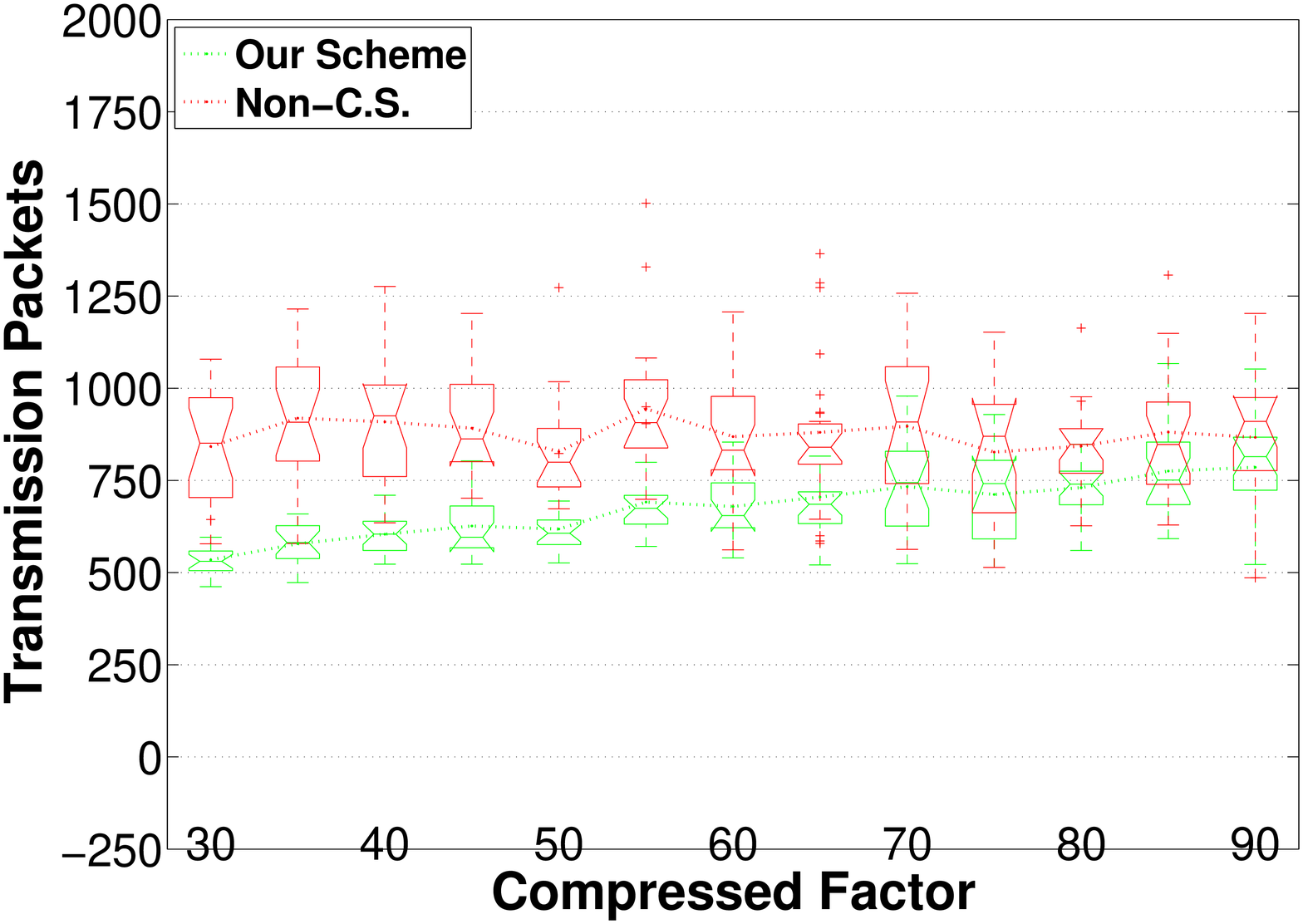}\label{Fig:tranComp}}
\subfigure[Cost Comparison-1024]{\includegraphics[width=4.2cm]{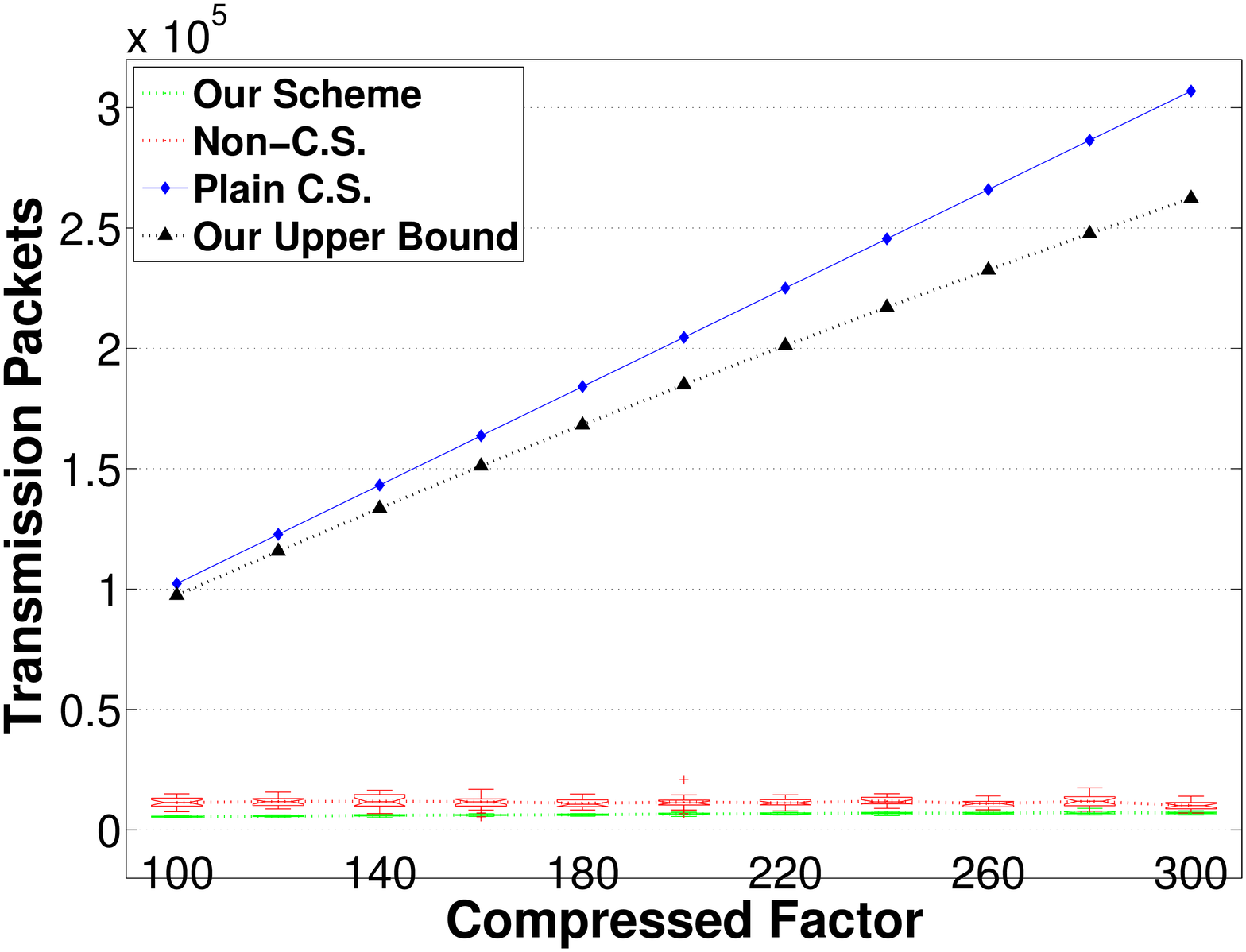}\label{Fig:tran1024}}
\subfigure[Comparison with Non-C.S.-1024]{\includegraphics[width=4.2cm]{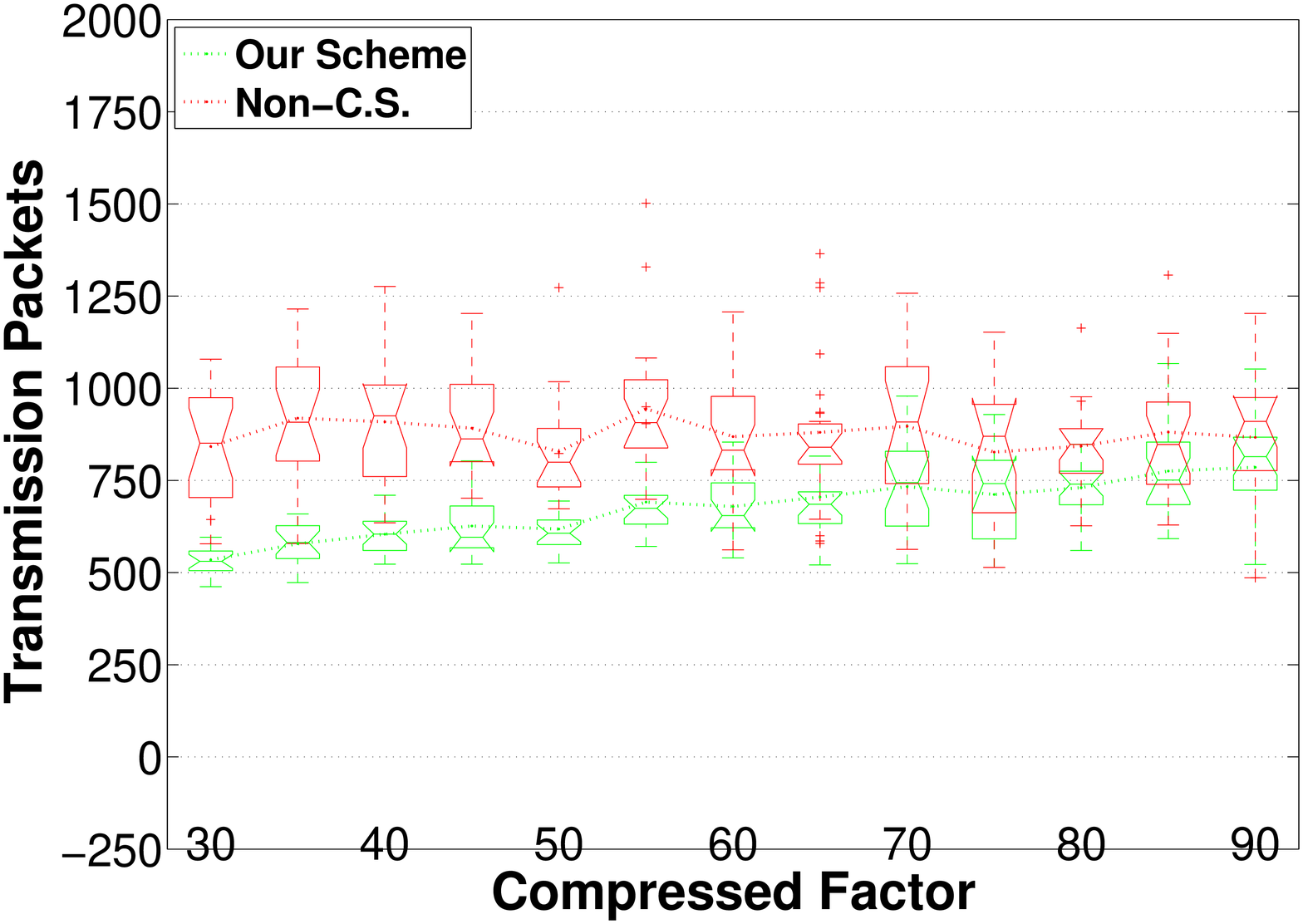}\label{Fig:tranComp2}}
\caption{Overall Transmission Comparison of 128 nodes and 1024 nodes\label{Fig:allRound}}
\end{figure*}

\begin{figure*}[ht!]
\centering
\subfigure[Performance of Round 2]{\includegraphics[width=4.2cm]{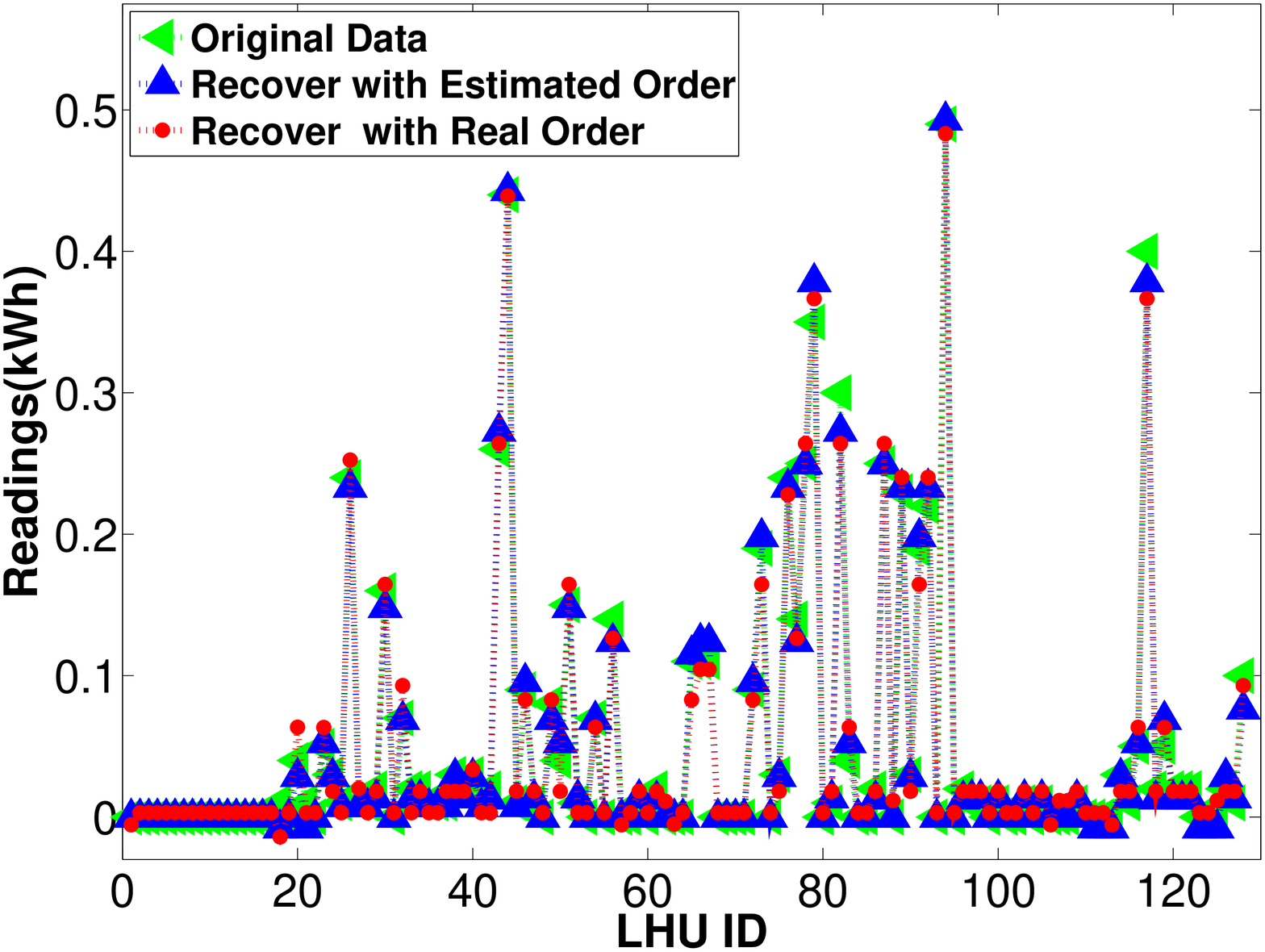}\label{Fig:round2}}
\subfigure[Performance of Round 66]{\includegraphics[width=4.2cm]{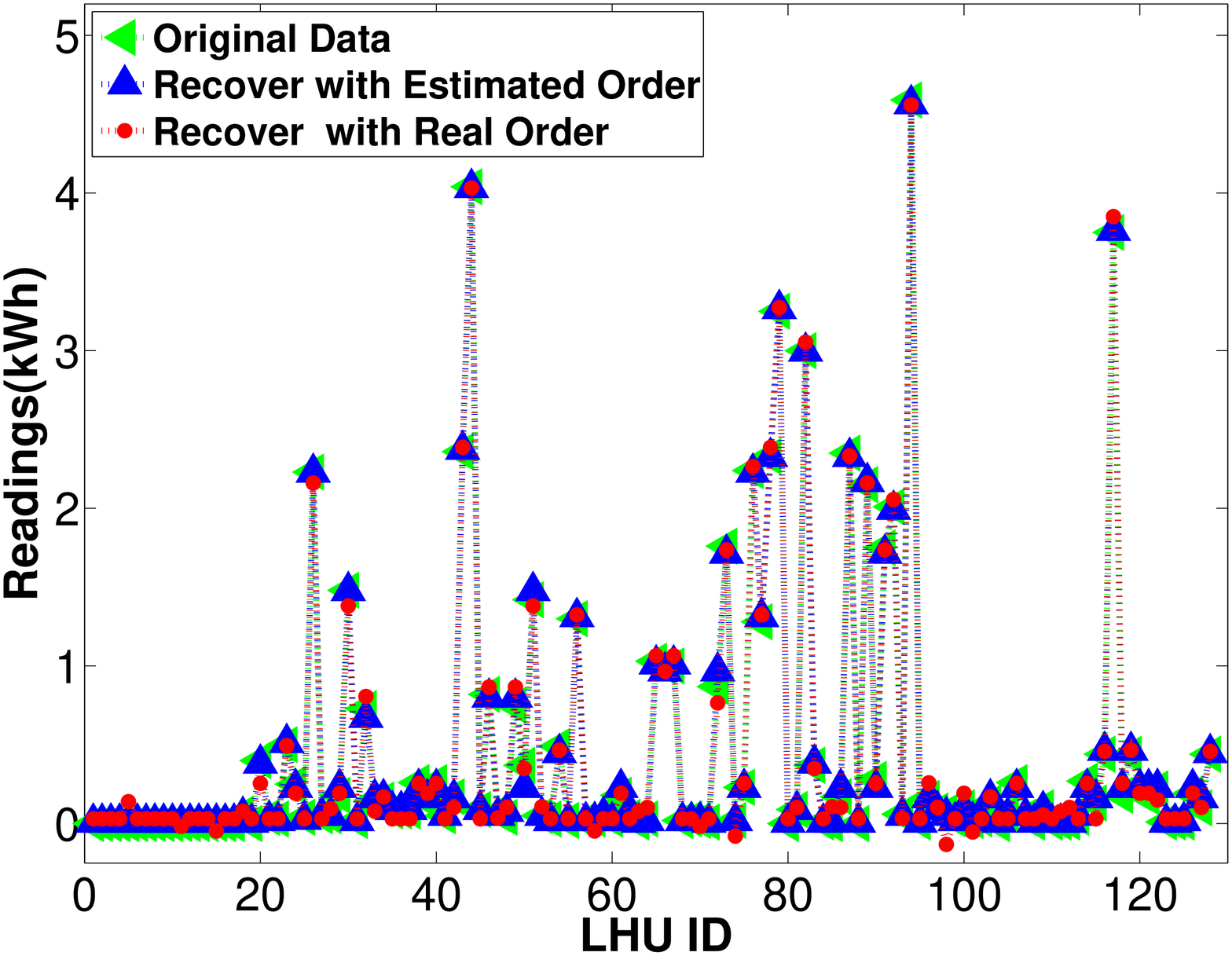}\label{Fig:round66}}
\subfigure[Performance of Round 132]{\includegraphics[width=4.2cm]{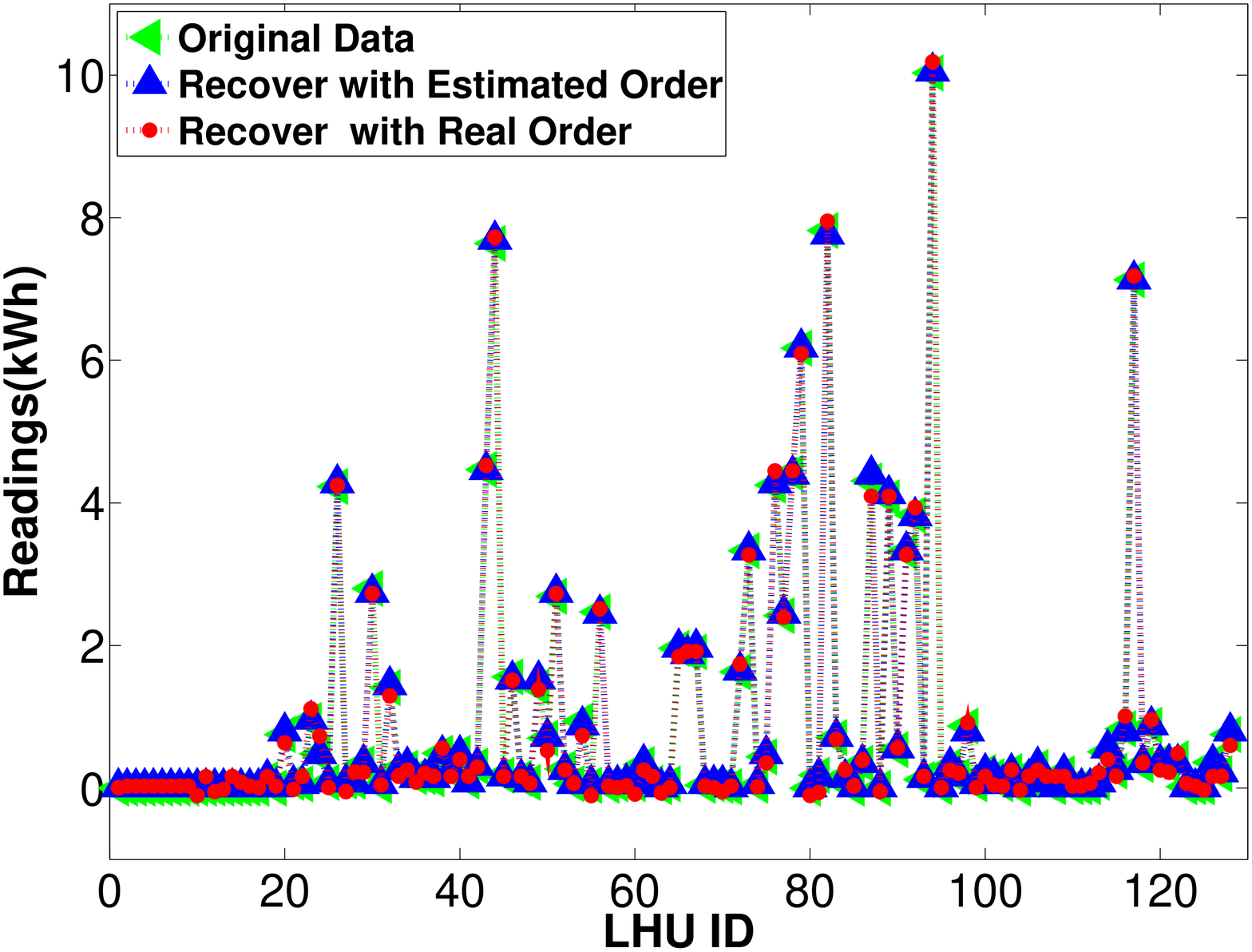}\label{Fig:round132}}
\subfigure[Performance of Round 198]{\includegraphics[width=4.2cm]{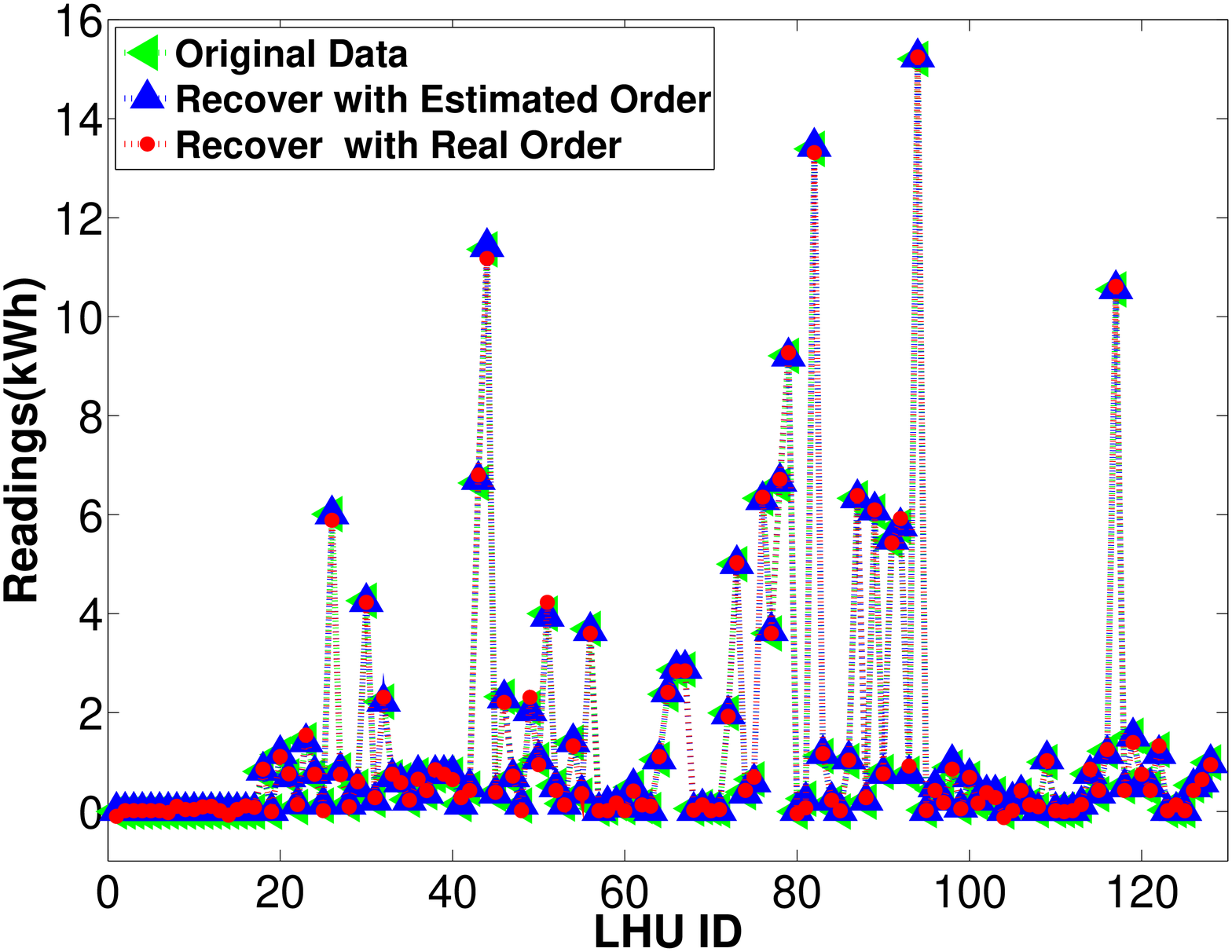}\label{Fig:round198}}
\subfigure[Performance of Round 264]{\includegraphics[width=4.2cm]{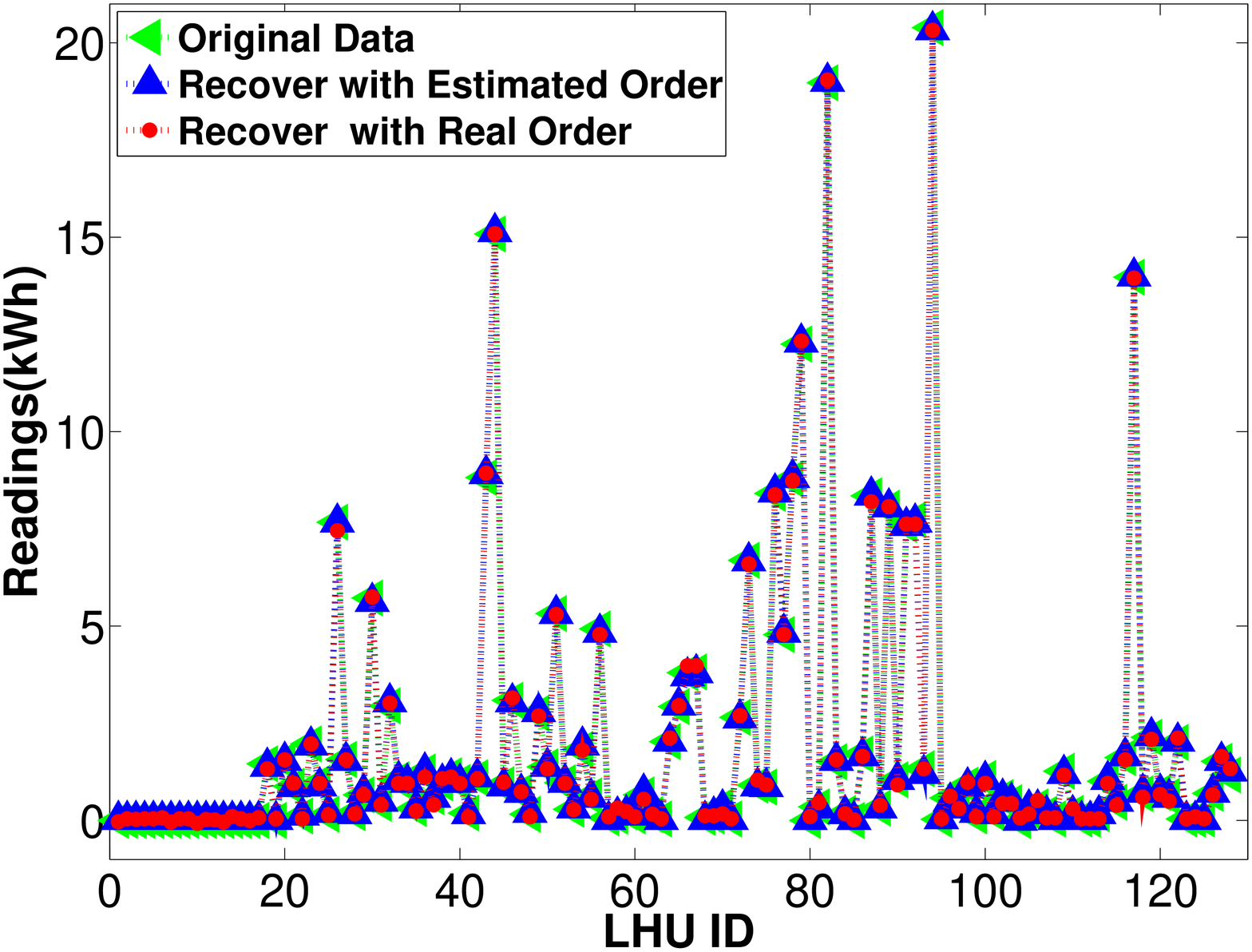}\label{Fig:round264}}
\subfigure[Performance of Round 330]{\includegraphics[width=4.2cm]{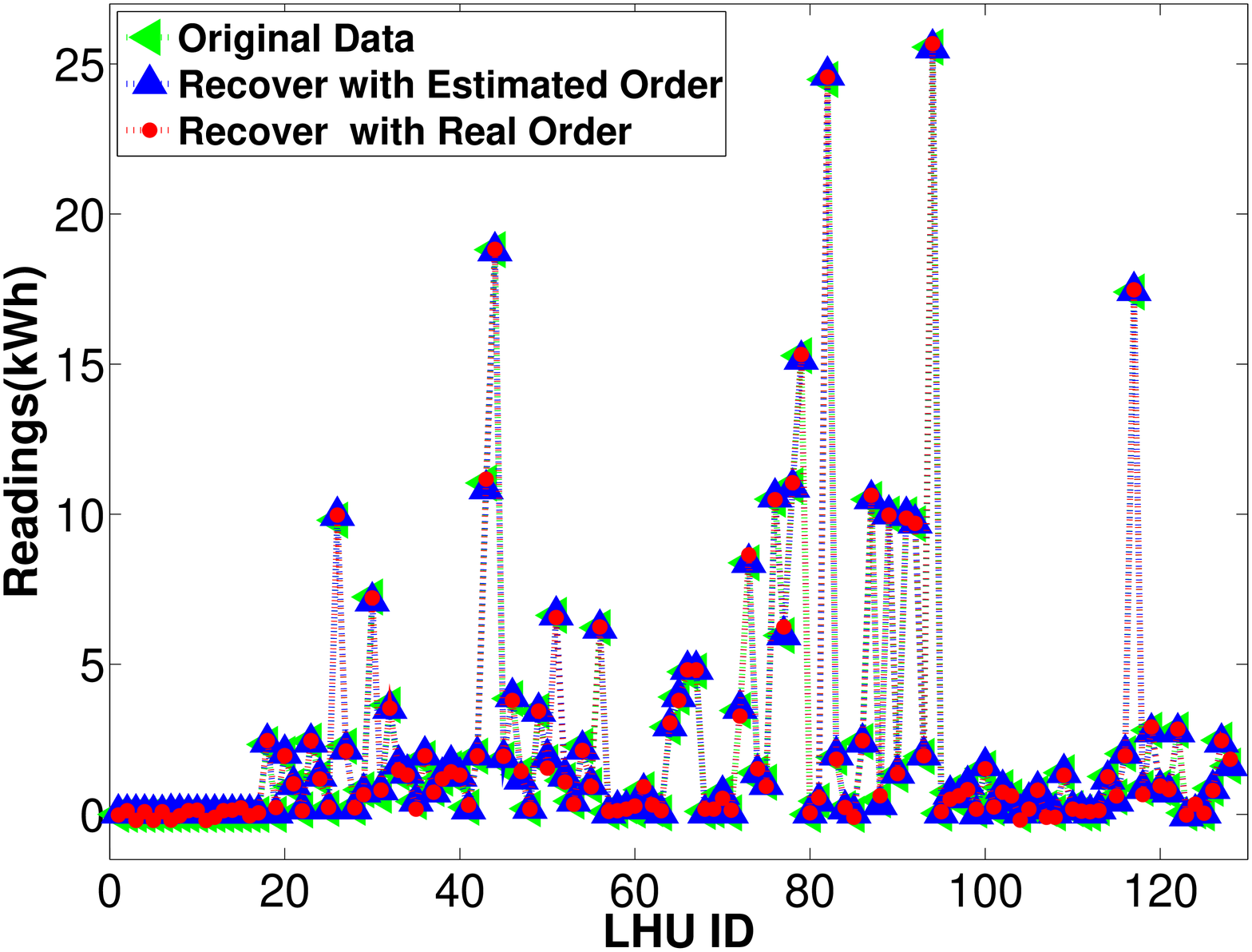}\label{Fig:round330}}
\subfigure[Performance of Round 396]{\includegraphics[width=4.2cm]{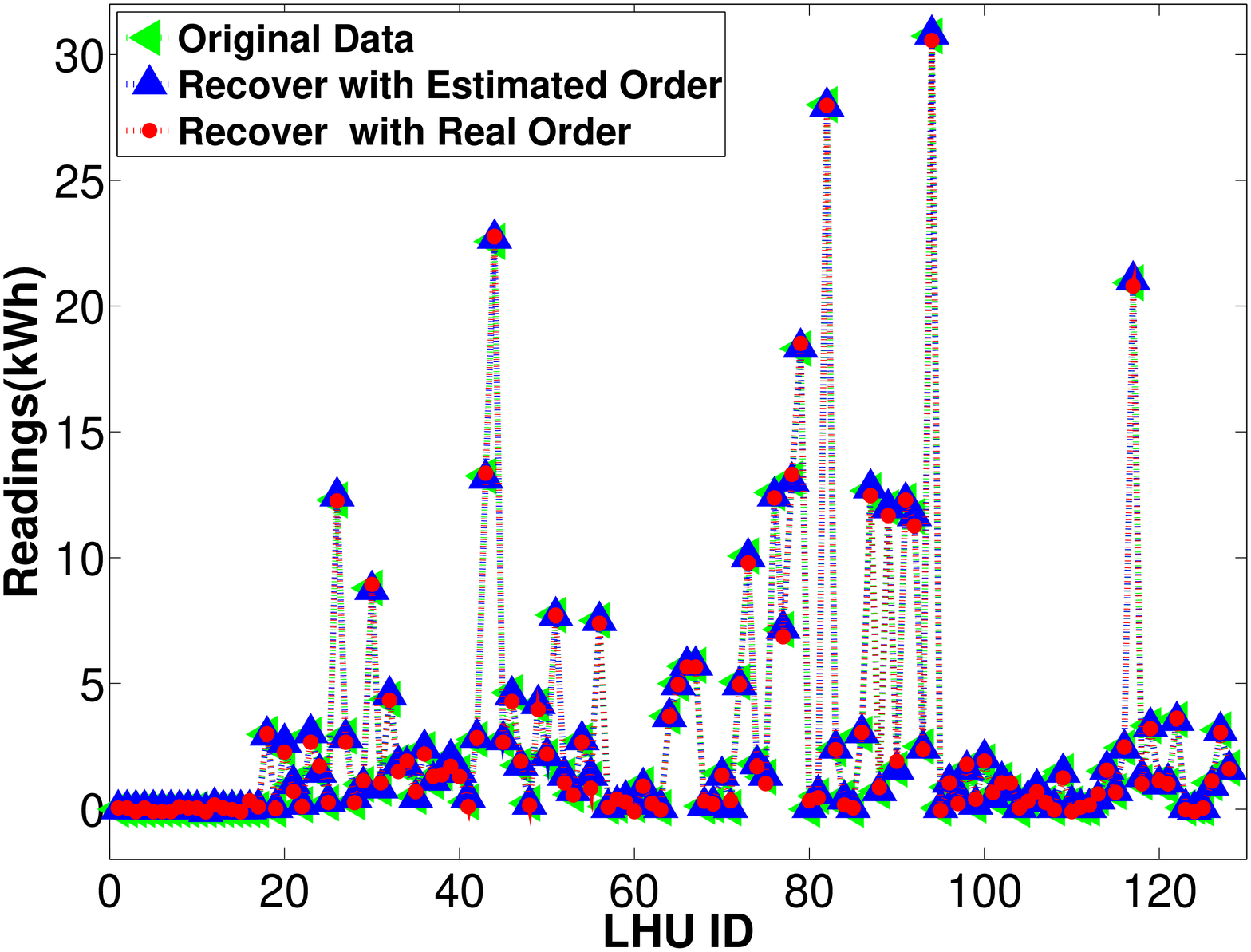}\label{Fig:round396}}
\subfigure[$L_{2}$ norm of Estimation Diiference]{\includegraphics[width=4.2cm]{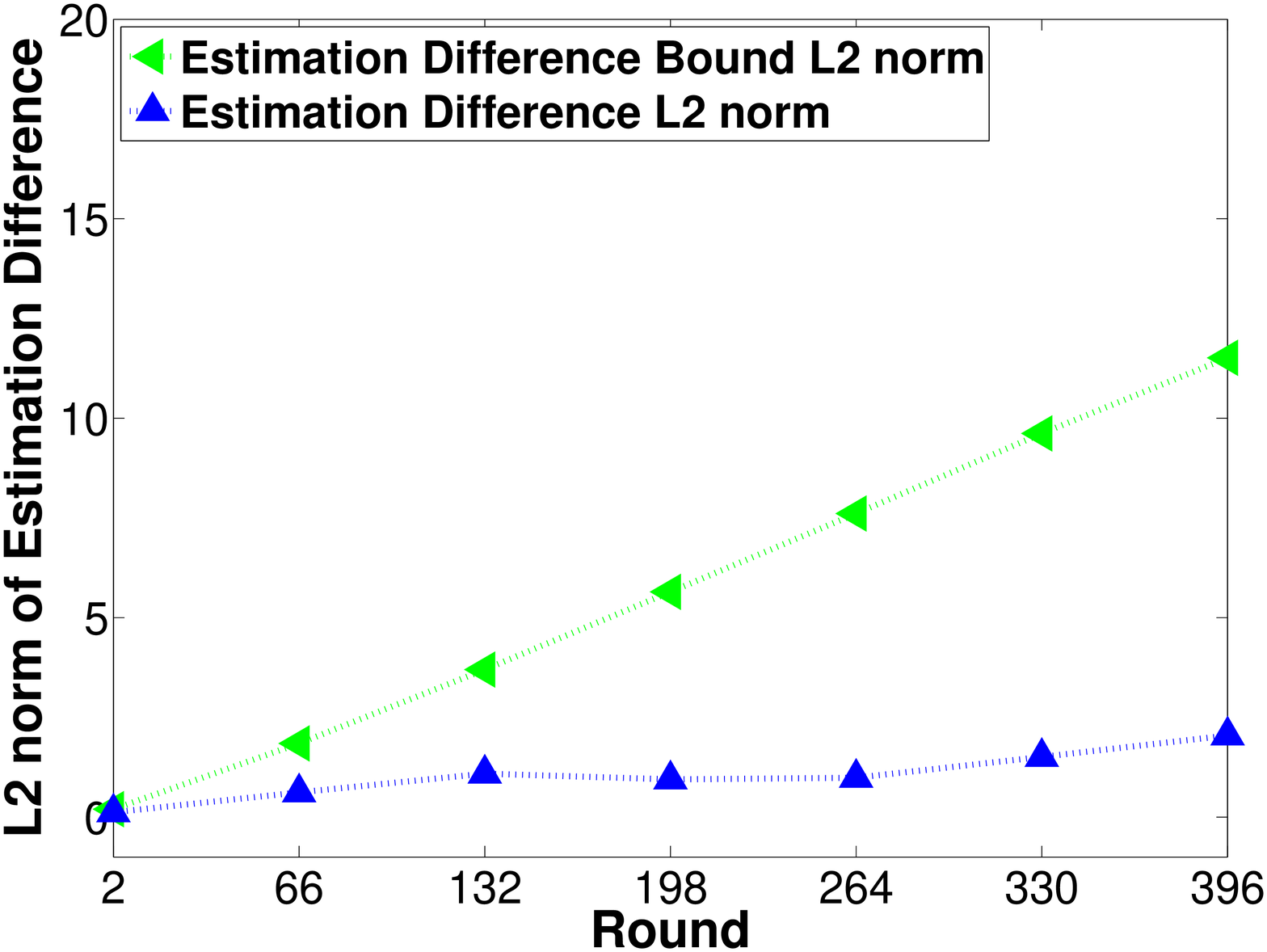}\label{Fig:estl2}}
\caption{Reconstruction Performance of Selected Transmission Rounds\label{Fig:allround}}
\end{figure*}

\par We first generate 20 arbitrary transmission networks as tree topology with 128 nodes and 1024 nodes respectively. Then assign each tree with specific compressed factor $M$, ranging around $0.3N$ where $N$ is the number of participated nodes; our companion work \cite{Mobicom09} has shown that $M=0.3N$ can achieve satisfactory recovery. Evaluation of transmission cost is performed over 20 different topologies given specific $M$ each time. We choose Box-plot to represent the information from 20 trees under specific $M$ and specific scheme.
\par We compare our cost with baseline scheme from \cite{Mobicom09} and transmission without Compressed Sensing aggregation. As Fig.\ref{Fig:tran128} and \ref{Fig:tran1024} show, since baseline scheme makes each node's outgoing link carry exact $M$ transmission packets, when $M$ and $N$ are given, transmission cost remained unchanged under different topologies and its box-plot is single value. Its overall transmission cost always exceeds our scheme upper bound. As for transmission without aggregation operations, we found it's more efficient than the referred baseline scheme. This result partly comes from the fact that tree topology spanning favors both width and depth instead of only length where topology more resembled the chain, and baseline scheme won more advantage.
\par Then compare our transmission cost with that of non-C.S. scheme. Fig.\ref{Fig:tranComp} and Fig.\ref{Fig:tranComp2} demonstrate the Box-plot of their results. For each $M$, our scheme reduce approximate $15\%$ transmission cost in comparison with non-C.S. under worst case; thus, it outperforms non-aggregation in general.

\subsubsection{Data Correlation analysis}\label{corranys}

\begin{figure}[ht!]
\centering
\subfigure[Correlation for Readings]{\includegraphics[width=4.2cm]{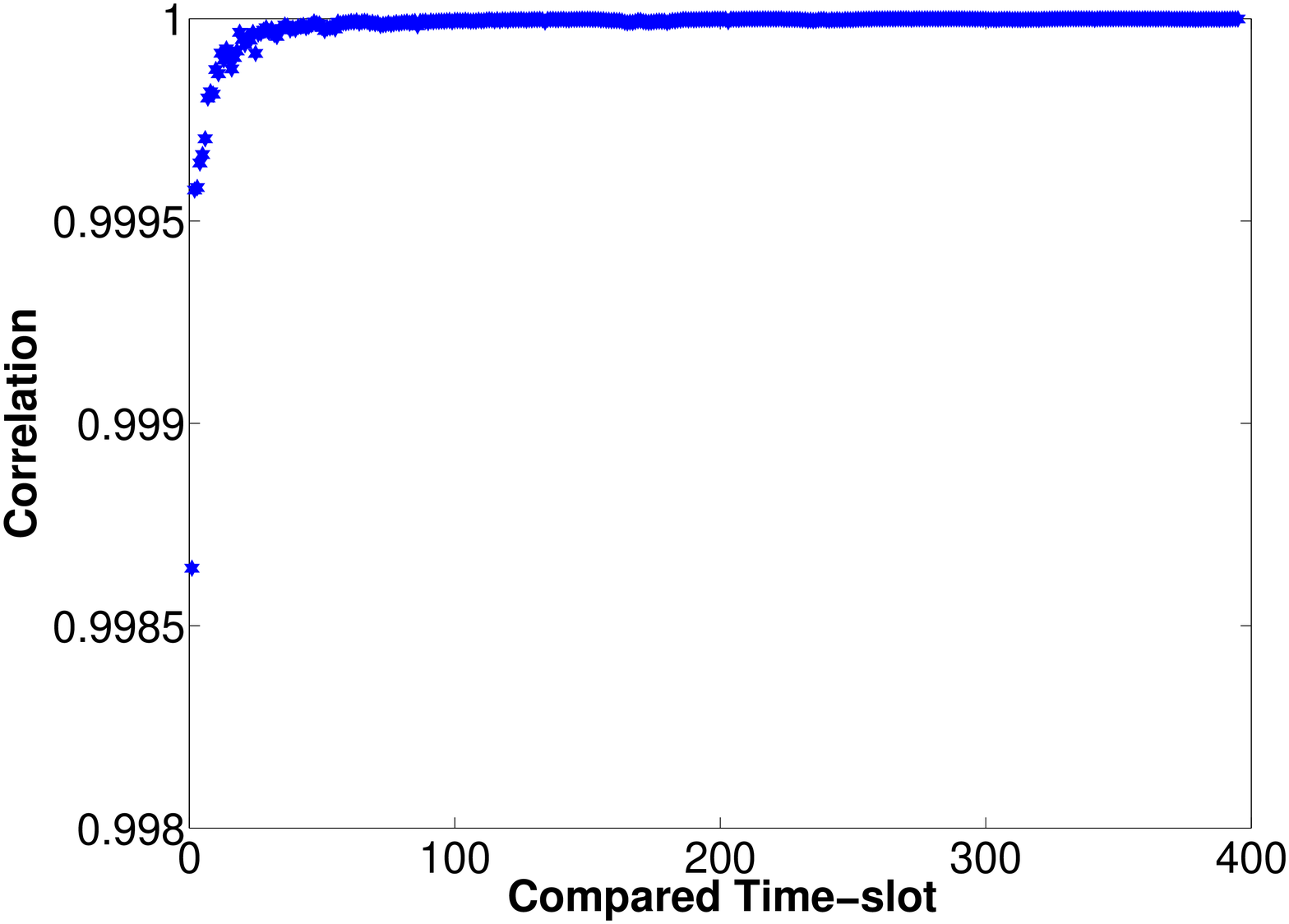}\label{Fig:datacorrelation}}
\subfigure[Correlation for Increment]{\includegraphics[width=4.2cm]{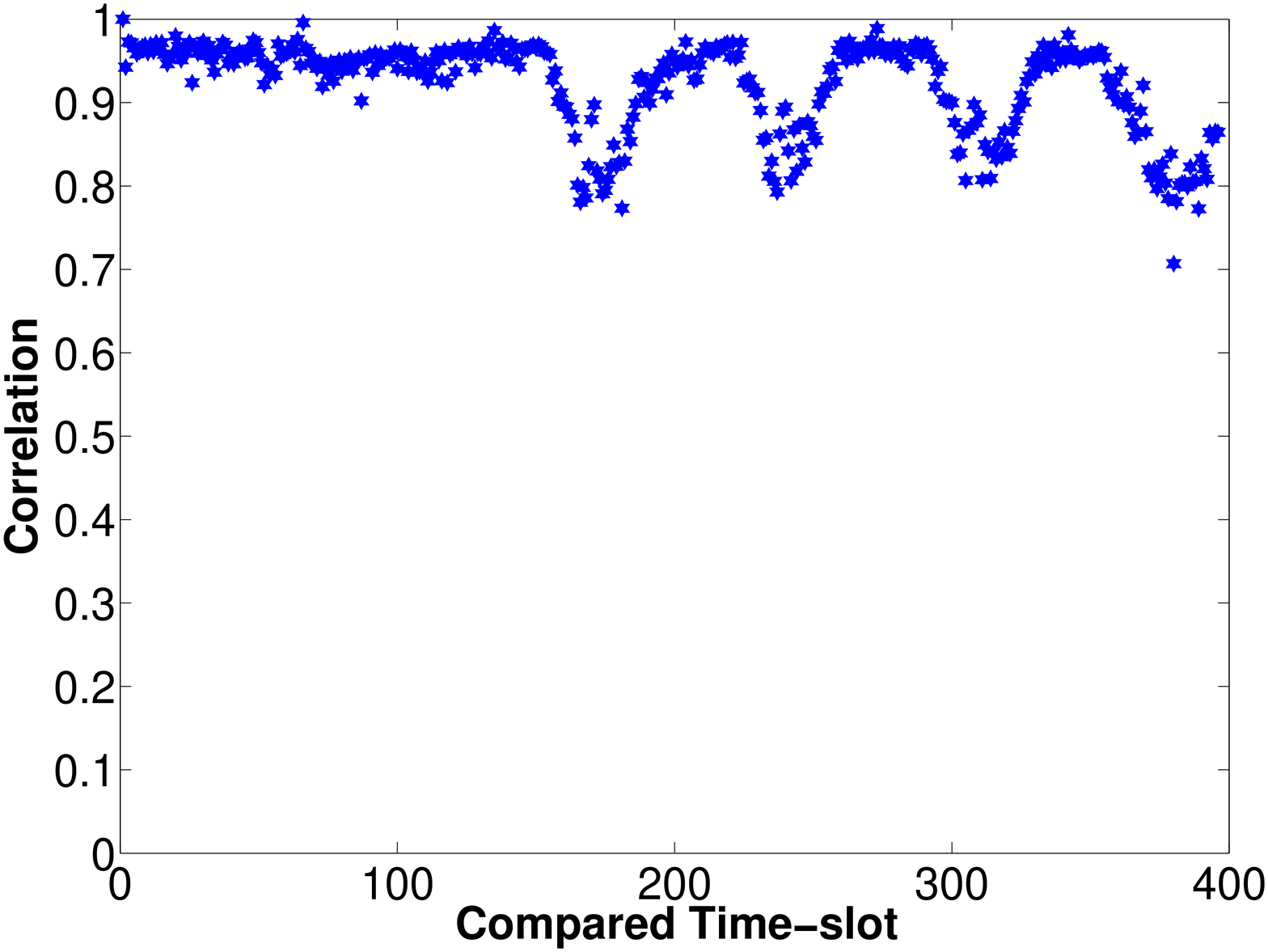}\label{Fig:noisecorrelation}}
\caption{Correlation for Increment and Readings\label{Fig:allround}}
\end{figure}
\par We make analysis against collected data and interval increment to reveal correlations.
\par First, we compare readings over 128 appliances at current time slot with the previous one from round 2 and get Fig.\ref{Fig:datacorrelation}. It represents strong correlations among the readings of every two time-slots, almost all correlations are larger than 99.95\%. This observation demonstrates that the order of readings at one time-slot wouldn't change too much in the next.
\par Then, we compare the readings at previous time slot and the increment between previous time slot and current one to get Fig.\ref{Fig:noisecorrelation}. We can observe that over 96.7\% correlations are larger than 0.8, which means the strong correlations among the readings and the increments. Therefore, the increments and readings have the similar patterns, thus share the same sparse domain.

\subsubsection{Reconstruction Performance}\label{recons}
\par Since our resorted readings are regarded as piecewise polynomial which can be represented with $K$-sparse under wavelet domain, and we have $N$ equal 128 from data collection, here we choose 7-level Haar wavelet domain for sparsity transform.
\par Here, we choose 7 rounds uniformly for evaluation. From Fig.\ref{Fig:round2} to Fig.\ref{Fig:round396}, we compare the reconstruction using our scheme with that choosing current readings' real order, also original readings are incorporated. First, we witness that the difference between reconstructions with our scheme using last round's estimated order and original readings don't amplify after 396 rounds transmissions, it can still achieve recovery up to acceptable bounded errors; moreover, difference between recovery with original data order and our scheme are quite small.
\par According to analysis from Section \ref{4.d}, when we use order of this round's estimated readings to reconstruct next ones, the errors incurred from choosing order with some inaccuracy would always be constrained within threshold of $C\beta_{A}\gamma_{A}||\boldsymbol{y}||_{2}$. Fig.\ref{Fig:estl2} describes the $L_{2}$ norm of estimation difference between measurements with estimated order and the original, for all 7 numerated rounds, our estimation errors are much less than the error bounds. The bound increases over rounds due to the increase of $l_{2}$ norm of compressed measurements $\boldsymbol{y}$; however, because of strong correlations in Fig.\ref{Fig:datacorrelation}, data sparsity remained well in DWT domain over multiple rounds and contribute to reconstruction performance.

\begin{figure}
\centering{}\includegraphics[width=0.3\textwidth]{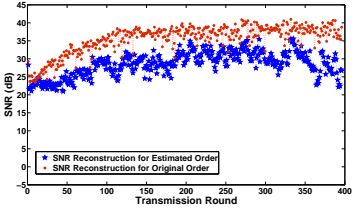}
\caption{SNR between Original Order and Estimated Order\label{SNRComp}}
\end{figure}

\par To further quantify the difference, we use SNR to evaluate the performance of our scheme: original readings' power as $P_{signal}$, power of the difference between reconstructed value and original as $P_{noise}$, we have $SNR$ = $10\lg(\frac{P_{signal}}{P_{noise}})$ and larger values represent better performance. From Fig.\ref{SNRComp}, for all 396 transmission rounds, SNRs of our scheme's reconstruction are generally less than that of reconstruction using original readings' order due to incurred errors from choosing some inaccurate order; however, we can still achieve SNR larger than 20 all the time.
\par Therefore, our scheme can achieve good performance even using last round's estimated order; the incurred errors wouldn't propagate and amplify across time since strong correlations exist across readings, little perturbation of order matrix wouldn't degrade our performance.

\subsection{Security Cost}
Due to dependable transmission scheme in Section \ref{sec}, the overall transmission cost would increase in packet size. The useful message length would increase 13 bytes from 6 bytes (4 for original readings, 2 for LHU ID) to 19 bytes (8 for encrypted readings, 2 for LHU ID, 9 for signature with hash and time-stamp). In practice, when considering other information required for transmission such as MAC and PHY Headers \cite{SGC10_SEC_Bartoli}, 13 bytes would only occupy small part of overall transmitted data and wouldn't degrade our transmission performance.

\section{Conclusion} \label{conclude}

\par In this paper, we have proposed a new scheme for efficient and secure meter reading in Smart Grids based on compressed sensing. This is the first attempt to solve the problems of efficiency, security and individual data transmission simultaneously. Our scheme works for collecting stream data, the incurred estimation errors wouldn't propagate over time.

\par We observe strong temporal correlations among real meter readings which indicate their sparsity in certain domain. Building upon this observation, we propose the compressed reading scheme which can work over arbitrary tree topologies and reduce total transmission cost that is needed to collect and recover all the meter readings. In contrast to traditional CS-based transmission scheme that requires to know the sparse domain a prior, our scheme can recover stream data \emph{without the knowledge of sparse domain beforehand}. We prove that the reconstruction error is bounded for every instance of the stream data and does not drift over time. Through numerical experiments, we observe that our scheme can reduce the transmission cost significantly as compared to a common benchmark \cite{Mobicom09} and the non-aggregation scheme. Due to the strong temporal correlations among the collected real-world data, we achieve good reconstruction performance. Meanwhile, we tailor specific security scheme to ensure reliability and security of data transmission. Moreover, it incurs only minor extra overhead. 
\section{Appendix}\label{app}
\subsection{Proof of Theorem \ref{errbnd}}
\begin{IEEEproof}
Original readings $\boldsymbol{d}=\Psi\boldsymbol{x}$, where $\boldsymbol{x}$ is the sparse representation of $\boldsymbol{d}$ in domain $\Psi$. Solution to (\ref{ptbCS}) is $\boldsymbol{\hat{x}}$. Since $H^{-1}$ is blind during reconstruction in (\ref{ptbCS}), we use $\Psi \boldsymbol{\hat{x}}$ as recovered readings $\boldsymbol{\hat{d}}$. Since wavelet matrix $\Psi$ is orthonormal, $||\boldsymbol{d}-\boldsymbol{\hat{d}}||_{l_{2}}$=$||\Psi(\boldsymbol{x}-\boldsymbol{\hat{x}})||_{l_{2}}$=$||\boldsymbol{x}-\boldsymbol{\hat{x}}||_{l_{2}}$
\par From definitions in (\ref{errdef}) and Theorem 2 in \cite{estbBound}, we can directly get (\ref{errcons}) to bound the reconstruction errors in our scheme with $\Psi$, $\Phi$, $\delta_{k}$, $\boldsymbol{y}$ and "worst" $H^{-1}$. Since the noise entry \textbf{\emph{e}} equals \textbf{0}, our perturbation only comes from matrix $H^{-1}$.
\end{IEEEproof}

\subsection{Proof of Proposition \ref{exact} and \ref{approximate}}
\begin{IEEEproof}
Since $\boldsymbol{y(t_{i+1})}=\Phi \boldsymbol{d(t_{i+1})}$ and $\boldsymbol{y(t_{i})}=\Phi \boldsymbol{d(t_{i})}$,
$\boldsymbol{y(t_{i+1})}-\boldsymbol{y(t_{i})}=\Phi[\boldsymbol{d(t_{i+1})}-\boldsymbol{d(t_{i})}]=\Phi \boldsymbol{n(t_{i})}$. Therefore,
the estimation error $||\boldsymbol{\hat{d}(t_{i+1})}-\boldsymbol{d(t_{i+1})}||_{l_{2}}=||\boldsymbol{d(t_{i})}+\boldsymbol{\hat{n}(t_{i})}-\boldsymbol{d(t_{i})}-\boldsymbol{n(t_{i})}||_{l_{2}}=||\boldsymbol{\hat{n}(t_{i})}-\boldsymbol{n(t_{i})}||_{l_{2}}$, $i=0,1,...,\infty$.
\par Denote $\boldsymbol{\hat{z}(t_{i})}$ and $\boldsymbol{z(t_{i})}$ are the representation of $\boldsymbol{\hat{n}(t_{i})}$ and $\boldsymbol{n(t_{i})}$ in domain $H^{-1}(t_{i})\Psi$, respectively. $||\boldsymbol{\hat{n}(t_{i})}-\boldsymbol{n(t_{i})}||_{l_{2}}=||H^{-1}(t_{i})\Psi\boldsymbol{\hat{z}(t_{i})}-H^{-1}(t_{i})\Psi\boldsymbol{z(t_{i})}||_{l_{2}}=||H^{-1}(t_{i})\Psi(\boldsymbol{\hat{z}(t_{i})}-\boldsymbol{z(t_{i})})||_{l_{2}}=||\boldsymbol{\hat{z}(t_{i})}-\boldsymbol{z(t_{i})}||_{l_{2}}$. From the Theorem 1.1 in \cite{candesrip}, $||\boldsymbol{\hat{z}(t_{i})}-\boldsymbol{z(t_{i})}||_{l_{2}}\leq C_{0}K^{-1/2}||\boldsymbol{z_{K}(t_{i})}-\boldsymbol{z(t_{i})}||_{l_{1}}$.
\begin{itemize}
    \item If $\boldsymbol{n(t_{i})}$ is $K$-sparse in domain $H^{-1}(t_{i})\Psi$ , then $||\boldsymbol{z_{K}(t_{i})}-\boldsymbol{z(t_{i})}||_{l_{1}}=0$. We can reconstruct $\boldsymbol{d(t_{i+1})}$ exactly.
    \item If $\boldsymbol{n(t_{i})}$ is approximately $K$-sparse in domain $H^{-1}(t_{i})\Psi$, $||\boldsymbol{\hat{d}(t_{i+1})}-\boldsymbol{d(t_{i+1})}||_{l_{2}}\leq C_{0}K^{-1/2}||\boldsymbol{z_{K}(t_{i})}-\boldsymbol{z(t_{i})}||_{l_{1}}\leq C_{0}K^{-1/2}\varepsilon$.
\end{itemize}
\end{IEEEproof}

\subsection{Proof of Theorem \ref{thmCost}}

\begin{figure}
  \subfigure[K Leafs-Transmission Topology]{
    \label{topology} 
    \begin{minipage}[b]{0.225\textwidth}
      \centering
      \includegraphics[width=1.15in]{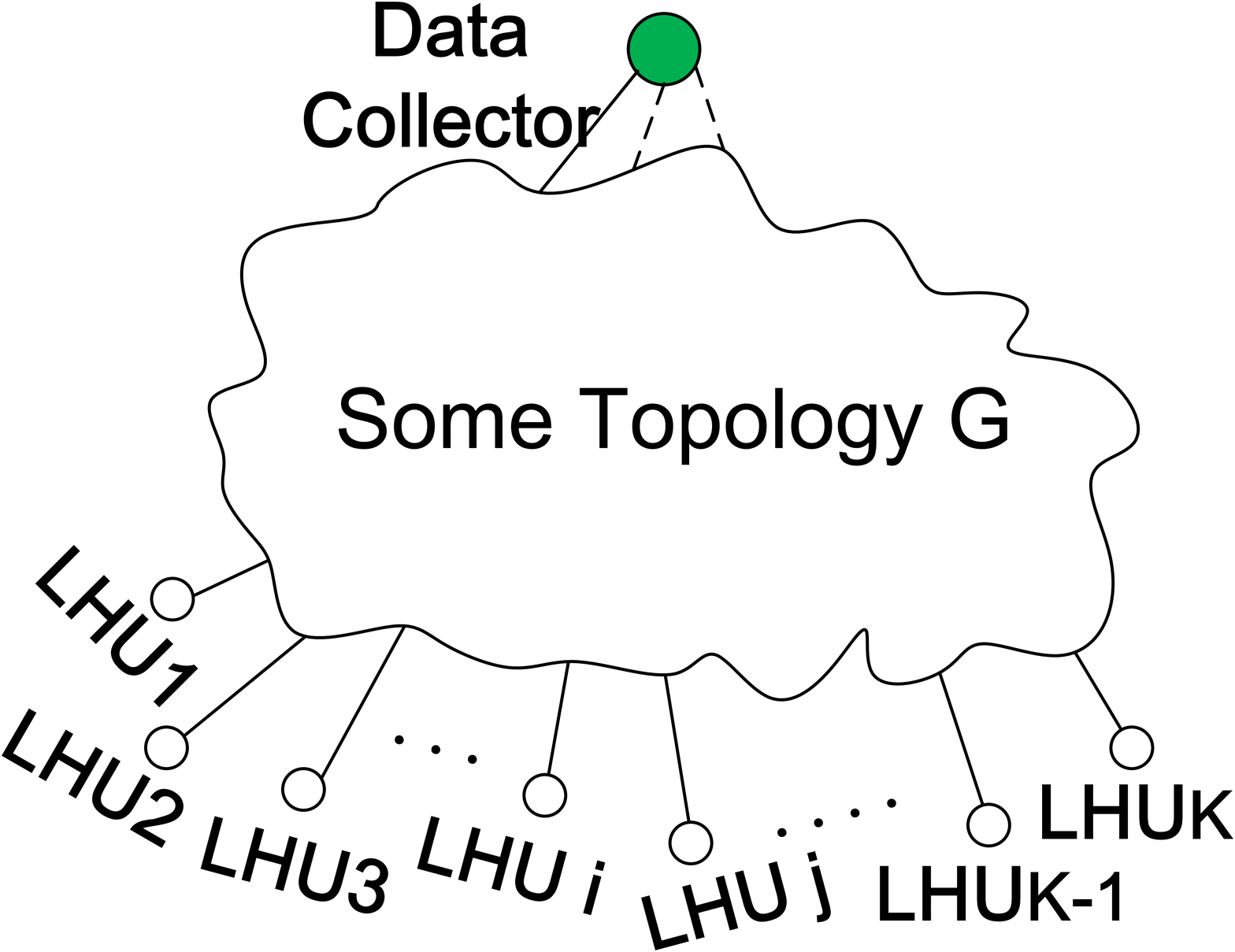}
    \end{minipage}}
  \subfigure[Merging two Nearest Branches]{
    \label{merge} 
    \begin{minipage}[b]{0.265\textwidth}
      \centering
      \includegraphics[width=1.5in]{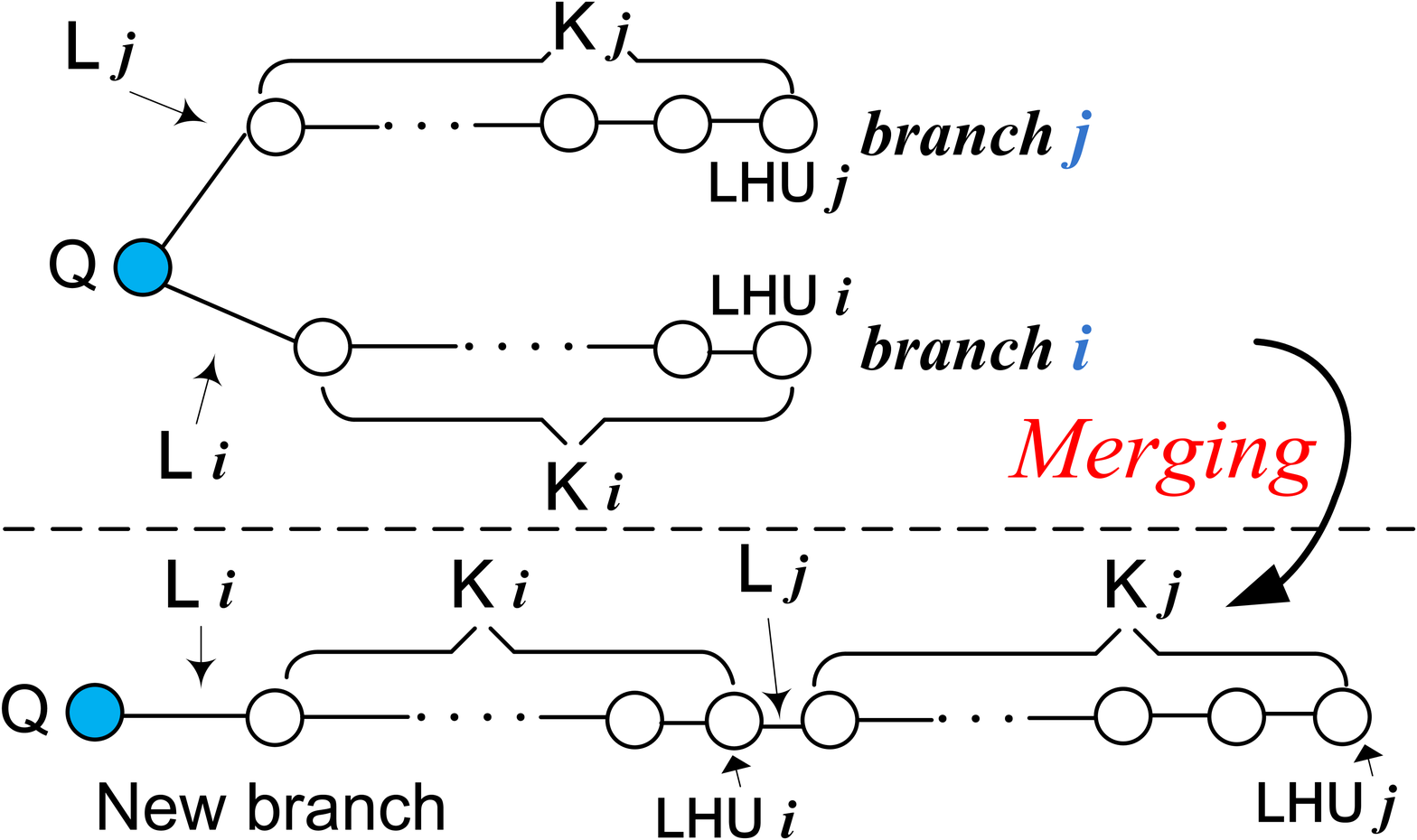}
    \end{minipage}}
\caption{Maximum Transmission Cost Proof}
  \label{fig:mini:subfig} 
\end{figure}
\begin{IEEEproof}
Under non-C.S. scheme, number of transmitted packets along each LHU's outgoing link $l_{i}$ ranges from 1 to M while always M when using CS strategy. According to this, overall transmitted packets $\sum_{i=1}^{N}l_{i}$ can vary from $N$ to $N$$\cdot$$M$ through respectively filling each link with 1 or $M$ packets.
\par Minimum cost $N$ can be achieved through broadcasting where all $N$ LHU directly transmits 1 packet towards collector. However, it's impossible to fill each link with $M$ packets due to assumption of tree topology where at least one leaf-LHU exists and its outgoing link carries only one packet.
\par Then discuss maximization case. Assume the cost can get maximized under topology with $K$ leaf-LHUs in Fig.\ref{topology}.

\par First, we choose two neighboring LHUs $i$ and $j$ and find nearest common ancestor (\textbf{NCE}) $Q$ as in Fig.\ref{merge}; thus, node $Q$ connects only two branches $i$ and $j$ as downstream children. Now we append branch $i$ to $j$'s tail and consider cost change.
\par Since property of LHU as aggregator or forwarder depends on its downstream nodes, property of $Q$ and its upstream nodes remain unchanged after appending. Thus consider change in branch $i$ and $j$ (including $L_{i}$ and $L_{j}$) while other transmission cost stays constant. Number of nodes in branch $i$ and $j$ are $K_{i}$ and $K_{j}$ respectively and we classify as following three cases.
\begin{itemize}
  \item $K_{i}, K_{j}$$\leq$$M$-1. After appending, transmission cost increases from $(K_{i}^{2}+K_{j}^{2}+K_{i}+K_{j})/2$ to $({(K_{i}+K_{j})}^{2}+K_{i}+K_{j})/2$
  \item $K_{i}\leq M-1$, $K_{j}\geq M$. After appending, cost changes from $K_{j}M-M^{2}/2+M/2+(K_{i}+1)K_{i}/2$ to $K_{j}M-M^{2}/2+M/2+K_{i}M$. Due to $K_{i}\leq M-1$, the cost also increases. Similar for $K_{i}\geq M$, $K_{j}\leq M-1$
  \item $K_{i}\geq M$, $K_{j}\geq M$. After appending, cost increases from $(K_{i}+K_{j})M-(M^{2}-M)$ to $(K_{i}+K_{j})M-(M^{2}-M)/2$.
\end{itemize}

\par When merging two neighboring branches under \textbf{NCE}, number of leaf LHUs decrease by 1 and transmission cost decreases as well. Iteratively performing this operation, overall transmission cost keeps decreasing till one leaf LHU left.
\par Maximum cost can be achieved under chain topology with only one leaf LHU and the cost is $M(N-M/2+1/2)$.

\end{IEEEproof}

\subsection{Proof of Proposition \ref{kary}}
\begin{IEEEproof}
Denote $N$ is the number of nodes except data collector in the network; $M$ is the number of packets transmitted for each aggregator; $L$
is the layer of the given tree and $TotalCost$ is the number of overall transmitted packets.

Obviously, there are $p^{i}$ nodes in the $i^{th}$ layer, thus $\sum_{j=0}^{L}p^{j}=N$.
Each node in the $i^{th}$ layer has
$\sum_{j=0}^{L-i}p^{j}-1$ children nodes. Assume $L-l^{th}$ layer is the first
layer where the nodes are aggregators, which means that $\sum_{j=0}^{l}p^{j}-1 \geq M$ and $\sum_{j=0}^{l-1}p^{j}-1<M$. To this end, the nodes in the $j^{th}$ layer, $j=1,2,...,L-l$ are aggregators and others are forwarders. $TotalCost$ of the scheme
is the combination of two parts: cost of forwarders and cost of aggregators.

\begin{eqnarray*}
TotalCost & = & \sum_{j=0}^{l-1}(p^{L-j}\sum_{i=0}^{j}p^{i})+M\sum_{j=1}^{L-l}p^{j}\\
 & = & \frac{1}{p-1}\sum_{j=0}^{l-1}p^{L-j}(p^{j+1}-1)\\
 & + & \frac{1}{p-1}Mp(p^{L-l}-1)\\
 & = & \frac{1}{p-1}(lp^{L+1}-\sum_{j=L-l+1}^{L}p^{j})+M\sum_{j=1}^{L-l}p^{j}\\
 & = & \frac{1}{p-1}[lp^{L+1}-\frac{1}{p-1}p^{L-l+1}(p^{l}-1)]\\
 & + & \frac{1}{p-1}Mp(p^{L-l}-1)\\
 & = & \frac{1}{p-1}[(lp^{L+1}-p^{L-l+1})+M(p^{L-l+1}-p)\\
 & = & N(l-p^{-l})+Mp^{-l}[N-1-(M-1)p]\\
 & = & O(N\log_{p}M)
\end{eqnarray*}
\end{IEEEproof}

\bibliographystyle{IEEEtran}
\bibliography{infocom12}

\begin{thebibliography}{10}
\providecommand{\url}[1]{#1}
\csname url@samestyle\endcsname
\providecommand{\newblock}{\relax}
\providecommand{\bibinfo}[2]{#2}
\providecommand{\BIBentrySTDinterwordspacing}{\spaceskip=0pt\relax}
\providecommand{\BIBentryALTinterwordstretchfactor}{4}
\providecommand{\BIBentryALTinterwordspacing}{\spaceskip=\fontdimen2\font plus
\BIBentryALTinterwordstretchfactor\fontdimen3\font minus
  \fontdimen4\font\relax}
\providecommand{\BIBforeignlanguage}[2]{{%
\expandafter\ifx\csname l@#1\endcsname\relax
\typeout{** WARNING: IEEEtran.bst: No hyphenation pattern has been}%
\typeout{** loaded for the language `#1'. Using the pattern for}%
\typeout{** the default language instead.}%
\else
\language=\csname l@#1\endcsname
\fi
#2}}
\providecommand{\BIBdecl}{\relax}
\BIBdecl

\bibitem{Obama}
\BIBentryALTinterwordspacing
Federal-Communications-Commission. (2010, Mar.) National broadband plan.
  [Online]. Available: \url{http://www.broadband.gov/download-plan/}
\BIBentrySTDinterwordspacing

\bibitem{SGload}
A.~Ricci, B.~Vinerba, E.~Smargiassi, I.~De~Munari, V.~Aisa, and P.~Ciampolini,
  ``Power-grid load balancing by using smart home appliances,'' in
  \emph{International Conference on Consumer Electronics, Digest of Technical
  Papers.}, 2008, pp. 1 --2.

\bibitem{AUSSG}
P.~Wolfs and S.~Isalm, ``Potential barriers to smart grid technology in
  australia,'' in \emph{Australasian Universities Power Engineering
  Conference}, Sept. 2009, pp. 1--6.

\bibitem{TMSG}
``T-mobile aiming for smarter grid,'' in \emph{Information Week}, Apr. 2009.

\bibitem{SGC10_SEC_Liu}
F.~Li, B.~Luo, and P.~Liu, ``Secure information aggregation for smart grids
  using homomorphic encryption,'' in \emph{First IEEE International Conference
  on Smart Grid Communications}, 2010, pp. 327 --332.

\bibitem{SMG_CS}
H.~Li, R.~Mao, L.~Lai, and R.~Qiu, ``Compressed meter reading for
  delay-sensitive and secure load report in smart grid,'' in \emph{First IEEE
  International Conference on Smart Grid Communications}, 2010.

\bibitem{SGC10_SEC_Bartoli}
A.~Bartoli, J.~Hern\'{a}ndez-Serrano, M.~Soriano, M.~Dohler, A.~Kountouris, and
  D.~Barthel, ``Secure lossless aggregation for smart grid m2m networks,'' in
  \emph{First IEEE International Conference on Smart Grid Communications},
  2010, pp. 333 --338.

\bibitem{CS_src}
D.~Donoho, ``Compressed sensing,'' \emph{Information Theory, IEEE Transactions
  on}, vol.~52, no.~4, pp. 1289 --1306, 2006.

\bibitem{Mobicom09}
C.~Luo, F.~Wu, J.~Sun, and C.~W. Chen, ``Compressive data gathering for
  large-scale wireless sensor networks,'' in \emph{Proceedings of the 15th
  annual international conference on Mobile computing and networking}, 2009,
  pp. 145--156.

\bibitem{ICC2010}
J.~Luo, L.~Xiang, and V.~V. Athanasios, ``Compressed data aggregation for
  energy efficient wireless sensor networks,'' in \emph{8th Annual IEEE
  Communications Society Conference on Sensor, Mesh, and Ad Hoc Communications
  and Networks}, June 2011.

\bibitem{SECON}
J.~Luo, L.~Xiang, and C.~Rosenberg, ``Does compressed sensing improve the
  throughput of wireless sensor networks?'' in \emph{2010 IEEE International
  Conference on Communications}, Salt Lake City, Utah, USA, May 2010.

\bibitem{General_Sec_1}
P.~McDaniel and S.~McLaughlin, ``Security and privacy challenges in the smart
  grid,'' \emph{IEEE Security and Privacy}, vol.~7, pp. 75--77, 2009.

\bibitem{General_Sec_2}
D.~Kundur, X.~Feng, S.~Liu, T.~Zourntos, and K.~Butler-Purry, ``Towards a
  framework for cyber attack impact analysis of the electric smart grid,'' in
  \emph{First IEEE International Conference on Smart Grid Communications},
  2010, pp. 244 --249.

\bibitem{Mallat}
S.~Mallat, \emph{A Wavelet Tour of Signal Processing, Third Edition: The Sparse
  Way}, 3rd~ed.\hskip 1em plus 0.5em minus 0.4em\relax Academic Press, 2008.

\bibitem{DLP}
E.~Candes and T.~Tao, ``Decoding by linear programming,'' \emph{Information
  Theory, IEEE Transactions on}, vol.~51, no.~12, pp. 4203 -- 4215, 2005.

\bibitem{SP}
R.~Baraniuk, M.~Davenport, R.~Devore, and M.~Wakin, ``A simple proof of the
  restricted isometry property for random matrices,'' \emph{Constr. Approx},
  vol. 2008, 2007.

\bibitem{walker}
J.~Walker, \emph{A primer on wavelets and their scientific applications}.\hskip
  1em plus 0.5em minus 0.4em\relax Chapman \& Hall/CRC, 1999.

\bibitem{videoCoding}
E.~Setton and B.~Girod, ``Video streaming with sp and si frames,'' in \emph{In
  Proc. Visual Commun. Image Proc}, 2005.

\bibitem{estbBound}
M.~Herman and T.~Strohmer, ``General deviants: An analysis of perturbations in
  compressed sensing,'' \emph{Selected Topics in Signal Processing, IEEE
  Journal of}, vol.~4, no.~2, pp. 342 --349, Apr. 2010.

\bibitem{CCS09}
Y.~Liu, P.~Ning, and M.~K. Reiter, ``False data injection attacks against state
  estimation in electric power grids,'' in \emph{Proceedings of 16th ACM
  Conference on Computer and Communications Security}, 2009.

\bibitem{SGC10_SE_1}
O.~Kosut, L.~Jia, R.~Thomas, and L.~Tong, ``Malicious data attacks on smart
  grid state estimation: Attack strategies and countermeasures,'' in
  \emph{First IEEE International Conference on Smart Grid Communications},
  2010, pp. 220 --225.

\bibitem{SGC10_SE_2}
G.~D\'{a}n and S.~Henrik, ``Stealth attacks and protection schemes for state
  estimators in power systems,'' in \emph{First IEEE International Conference
  on Smart Grid Communications}, 2010, pp. 214 --219.

\bibitem{SGC10_SEC_Scheme}
G.~Kalogridis, C.~Efthymiou, Z.~D. Stojan, A.~L. Tim, and R.~Cepeda, ``Privacy
  for smart meters: Towards undetectable appliance load signatures,'' in
  \emph{First IEEE International Conference on Smart Grid Communications},
  2010, pp. 232 --237.

\bibitem{Paillier}
P.~Paillier, ``Public-key cryptosystems based on composite degree residuosity
  classes,'' in \emph{Proceedings of the 17th international conference on
  Theory and application of cryptographic techniques}, Berlin, Germany, 1999,
  pp. 223--238.

\bibitem{boxplot}
J.~W.~T. Robert~McGill and W.~A. Larsen, ``Variations of box plots,'' \emph{The
  American Statistician}, vol.~32, pp. 12--16, Feb. 1978.

\bibitem{PowerNet}
M.~Kaz, O.~Gnawali, O.~Heller, P.~Levis, and C.~Kozyrakis, ``Identifying energy
  waste through dense power sensing and utilization monitoring,'' in
  \emph{Technical Report}, Stanford University, 2010.

\bibitem{Tex15}
T.~C. Knisley, ``Advanced metering system: Conquering the challenge of mass
  deployment,'' Oncor Electric Delivery, May 2010.

\bibitem{candesrip}
E.~CANDES, ``The restricted isometry property and its implications for
  compressed sensing,'' \emph{Comptes Rendus Mathematique}, vol. 346, pp.
  589--592, 2008.

\end{thebibliography}
\end{document}